\documentclass[12pt]{article}

 \usepackage{amssymb}
\newcommand{\E}{\mathbb{E}}
\newcommand{\T}{\mathbb{T}}
\newcommand{\R}{\mathbb{R}}
\newcommand{\Z}{\mathbb{Z}}
\newcommand{\N}{\mathbb{N}}
\newcommand{\CC}{\mathbb{C}}

\newcommand{\beqn}{\begin{eqnarray}}
\newcommand{\eeqn}{\end{eqnarray}}
\newcommand{\be}{\begin{equation}}
\newcommand{\ee}{\end{equation}}
\newcommand{\ba}{\begin{array}}
\newcommand{\ea}{\end{array}}

\newcommand{\pa}{\partial}
\newcommand{\fr}{\frac}
\newcommand{\ov}{\overline}
\newcommand{\ds}{\displaystyle}
\newcommand{\ve}{\varepsilon}

\newcommand{\PV}{{\rm PV}}

\newcommand{\tg}{\mathop{\rm tg}\nolimits}
\newcommand{\const}{\mathop{\rm const}\nolimits}
\newcommand{\sgn}{\mathop{\rm sign}\nolimits}
\newcommand{\tr}{\mathop{\rm tr}\nolimits}
\newcommand{\supp}{\mathop{\rm supp}\nolimits}

\newcommand{\bo}{{\hfill\loota}}
\newcommand{\loota}{\hbox{\enspace{\vrule height 7pt depth 0pt width 7pt}}}

\begin{document}

\renewcommand{\theequation}{\thesection.\arabic{equation}}
\newtheorem{theorem}{Theorem}[section]
\renewcommand{\thetheorem}{\arabic{section}.\arabic{theorem}}
\newtheorem{definition}[theorem]{Definition}
\newtheorem{deflem}[theorem]{Definition and Lemma}
\newtheorem{lemma}[theorem]{Lemma}
\newtheorem{example}[theorem]{Example}
\newtheorem{examples}[theorem]{Examples}
\newtheorem{remark}[theorem]{Remark}
\newtheorem{remarks}[theorem]{Remarks}
\newtheorem{cor}[theorem]{Corollary}
\newtheorem{pro}[theorem]{Proposition}

\begin{titlepage}

\begin{center}
 {\Large\bf  On the energy current for harmonic crystals}\\
 \vspace{2cm}
{\large T.V.~Dudnikova}\medskip\\
{\it  M.V.Keldysh Institute of Applied Mathematics RAS\\
 Moscow 125047, Russia}\medskip\\
e-mail:~ tdudnikov@mail.ru
 \end{center}
 \vspace{1cm}

\begin{abstract}
We consider a $d$-dimensional harmonic crystal,  $d \ge 1$,
and study the Cauchy problem with random initial data.
We assume that the random initial function is
close to different translation-invariant
processes for large values  of $x_1,\dots,x_k$ with some $k\in\{1,\dots,d\}$.
The distribution $\mu_t$  of the  solution at time $t\in\R$ is studied.
We prove the convergence of correlation functions of the measures $\mu_t$
to a limit  for large times.
The explicit formulas  for the limiting
correlation functions and for the energy current density (in mean)
are obtained in the terms of the initial covariance.
We give the application to the case of the Gibbs initial measures  with  different temperatures.
In particular, we find stationary states in which there is a constant non-zero energy current flowing through
the harmonic crystal.
Furthermore, the weak convergence of $\mu_t$ to a limit measure is proved.
We also study the initial boundary value problem for the harmonic crystal
with zero boundary condition and obtain the similar results.
 \medskip

\medskip

 \end{abstract}
\end{titlepage}

\section{Introduction}

We study the Cauchy problem for
a harmonic crystal in $d$ dimensions with $n$ components,  $d,n \ge 1$.
We assume that the initial datum $Y_0(x)$, $x=(x_1,\dots,x_d)\in\Z^d$, of the problem is a
random element of the Hilbert space ${\cal H}_\alpha$ consisting of real sequences,
see Definition~\ref{d1.1} below.
  The distribution of $Y_0(x)$ is a probability measure  $\mu_0$
with zero mean value. We assume that the covariance $Q_0(x,y)$ of $\mu_0$
decreases like $|x-y|^{-N}$ as $|x-y|\to\infty$ with some $N>d$. Furthermore,
we impose the condition {\bf S3} (see formulas (\ref{3.1})--(\ref{3a}) below) which
 means roughly that
$Y_0(x)$ is close to different translation-invariant processes $Y_{\bf n}(x)$
with distributions $\mu_{\bf n}$  as $(-1)^{n_j}x_j\to+\infty$ for all $j=1,\dots,k$,
 with some $k\in\{1,\dots,d\}$. Here ${\bf n}$ stands for the vector ${\bf n}=(n_1,\dots, n_k)$,
where all $n_j\in\{1,2\}$.
Given $t\in\R$, denote by $\mu_t$ the probability measure
that gives the distribution of the  solution $Y(x,t)$ to dynamical equations
with the random initial datum $Y_0$.
 We study the asymptotics of $\mu_t$ as $t\to\infty$.
The first objective is
to prove the convergence of the correlation functions of $\mu_t$ to a limit,
 \be\label{corf}
Q_t(x,y)\equiv\ds\int_{{\cal H}_\alpha} \Big(Y_0(x)\otimes Y_0(y)\Big)\, \mu_t(dY_0)
\to Q_\infty(x,y),\quad t\to\infty,\quad x,y\in\Z^d.
\ee
The explicit formulas for the limit covariance $Q_\infty$ are  given in
(\ref{1.13})--(\ref{C(theta)}).
They allow us to derive the expression for the limiting
 mean energy current density ${\bf J}_\infty$ in the terms of the initial covariance $Q_0(x,y)$.
\medskip

We apply our results to a particular case when $\mu_{{\bf n}}$  are Gibbs measures
with different  temperatures  $T_{\bf n}>0$. 
Therefore, our model can be considered as a ``system + $2^k$ reservoirs'',
where ``reservoirs'' consist of the crystal particles lying in $2^k$ regions  of a form
$\{x\in\Z^d:\,(-1)^{n_j}x_j>a\,\,\,\mbox{for all }\,j=1,\dots,k,\,\,\mbox{where }\,n_j=1\,\mbox{or }\,2\}$
with some $a>0$,
 and the ``system'' is the remaining part of the crystal.
 At $t=0$, the reservoirs have Gibbs distributions with corresponding temperatures
 $T_{\bf n}$, ${\bf n}=(n_1,\dots,n_k)$.
 (In the case of $d=1$, the similar model was studied  by Spohn and Lebowitz \cite{SL}.)
We show that the energy current density ${\bf J}_\infty$ is a constant vector
 satisfying formulas (\ref{mecd}) and (\ref{4.11}).
Furthermore, under additional symmetry conditions on the harmonic crystal,
 the coordinates of the energy current ${\bf J}_\infty\equiv (J^1_\infty,\dots,J^d_\infty)$
are of a form
\be\label{Jl}
J^l_\infty=\left\{\ba{ll}
-c_l\sum'\Big(T_{\bf n}\Big|_{n_l=2}-T_{\bf n}\Big|_{n_l=1}\Big)
& \mbox{for }\,\quad l=1,\dots,k,\\
0& \mbox{for }\,\quad l=k+1,\dots,d,\ea\right.\ee
with some constants $c_l>0$. Here  the summation $\sum'$ is taken over
all $n_j$ with $j\not=l$ .
\medskip

Our second result gives  the (weak) convergence of the measures
$\mu_t$ on the Hilbert space ${\cal H}_\alpha$
with $\alpha<-d/2$ to a limit measure $\mu_{\infty}$,
\be\label{1.8i}
\mu_t \rightharpoondown \mu_\infty,\quad t\to \infty.
\ee
This means the convergence of the integrals
$$
 \int f(Y)\mu_t(dY)\rightarrow  \int f(Y)\mu_\infty(dY) \quad \mbox{as }\,\,\, t\to \infty,
$$
 for any bounded continuous functional $f$  on ${\cal H}_\alpha$.
Furthermore, the  limit measure $\mu_{\infty}$
 is a translation-invariant Gaussian measure on ${\cal H}_\alpha$    
and has the mixing property.
\smallskip

For infinite one-dimensional (1D) chains of harmonic oscillators,
similar results  have  been established by Boldrighini, Pellegrinotti and Triolo \cite{BPT}
and by Spohn and Lebowitz \cite{SL}.
In earlier investigations, Lebowitz {\em et al.} \cite{RLL, CaLeb},
 Nakazawa \cite{Na} analyzed the stationary energy current through
the finite 1D chain of harmonic oscillators in contact with
external heat reservoirs at different temperatures.
For $d\ge 1$, the convergence (\ref{1.8i}) has been obtained
for the first time  by Lanford and  Lebowitz \cite{LL} for initial measures which are absolutely
continuous with respect to the canonical Gaussian measure.
We consider more general class of
 initial measures with the mixing condition and do not assume the absolute
continuity. For the first time the mixing condition has been introduced by Dobrushin and
Suhov for the ideal gas \cite{DS}. Using the mixing condition,
we have proved the convergence for the wave and Klein--Gordon equations
(see \cite{DKM2} and references therein)
 for non translation invariant  initial measures $\mu_0$.
For many-dimensional crystals, the results (\ref{corf}) and (\ref{1.8i})
were obtained in \cite{DKS1} for translation invariant  measures $\mu_0$.
The present paper develops our previous work \cite{DKM1},
where (\ref{corf})--(\ref{1.8i}) were proved in the case of $k=1$.

In this paper, we also study the initial boundary value problem for
the harmonic crystal in the half-space $\Z^d_+=\{x\in\Z^d:x_1\ge0\}$
with {\em zero} boundary condition (as $x_1=0$)
and obtain the results similar to (\ref{corf}) and (\ref{1.8i}).
This generalizes the results of \cite{D08} on the more general class of the initial measures.
For this model, we calculate the limiting energy current density ${\bf J}_{+,\infty}(x_1)$,
see formulas (\ref{mecd+})--(\ref{7.17}).
In particular, if $d=1$, then   ${\bf J}_{+,\infty}(x_1)\equiv0$.
For any $d\ge2$, ${\bf J}_{+,\infty}(0)=0$.
For $d\ge2$ and $x_1>0$, the coordinates  of  ${\bf J}_{+,\infty}(x_1)$
have a form similar to (\ref{Jl}), but with  positive functions $c_l=c_l(x_1)$ if $l=2,\dots,k$,
and vanish if $l=1,k+1,\dots,d$.
Furthermore, ${\bf J}_{+,\infty}(x_1)$ tends to a limit as $x_1\to+\infty$ (see formula (\ref{7.18})).
For the 1D infinite chain of harmonic oscillators on the half-line with {\em nonzero} boundary
condition, we prove the results (\ref{corf}) and (\ref{1.8i})  in \cite{D17}
and show that there is a negative limiting energy current at origin (see \cite[Remark 2.11]{D17}).
\medskip

There are  a large literature devoted to the study of  return to equilibrium,
convergence to non-equilibrium states and heat conduction for nonlinear  systems, see
\cite{BLR, Lepri} for an extensive list of references.
For instance, ergodic properties and long time behavior were studied
for weak perturbation  of the infinite chain of harmonic oscillators
as a model of 1D harmonic crystals with defects by Fidaleo and Liverani \cite{FL} and  for
the finite chain of anharmonic oscillators coupled to a single heat bath by Jak\v{s}i\'c and Pillet \cite{JP}.
A finite chain of nonlinear oscillators coupled
to two heat reservoirs has been studied by Eckmann, Rey-Bellet and others \cite{EPR1, EPR2, RT}.
For such system the existence of non-equilibrium states
 and convergence to them have been investigated in \cite{EPR1, RT}.
In \cite{EPR2}, Eckmann, Pillet, and Rey-Bellet showed that heat (in mean)
flows  from the hot reservoir to the cold one. 
Fourier's law for a harmonic crystal with stochastic reservoirs was
proved by Bonetto, Lebowitz and  Lukkarinen \cite{BLL}.
In the present paper,
we find stationary non-equilibrium states 
in which there is a non-zero
energy current flowing through the {\it infinite} $d$-dimensional  harmonic crystal.
\medskip

 The paper is organized as follows.
 In Sec.~\ref{sec2}, we impose the conditions on the model and on
the initial measures $\mu_0$ and state the main results.
In Sec.~\ref{sec3}, we construct   examples of
 random initial data  satisfying all assumptions imposed.
The application to Gibbs initial measures and the derivation
of the formula (\ref{Jl}) are given in Sec.~\ref{sec4}.
In Sec.~\ref{sec5}, the uniform bounds for covariance of $\mu_t$ are obtained,
and  the proof of (\ref{1.8i}) is discussed.
 The asymptotics (\ref{corf}) is proved in Sec.~\ref{sec5.2}.
In Sec.~\ref{sec7}, we study
the initial-boundary value problem for harmonic crystals in the half-space
and prove the results similar to (\ref{corf})--(\ref{1.8i}).

\setcounter{equation}{0}
 \section{Main results}\label{sec2}
\subsection{Model}\label{sec2.1}
We consider a Bravais lattice in $\R^d$ with a unit cell which contains a
finite number of atoms. For notational simplicity, the lattice is assumed to be simple hypercubic.
Let
$u(x)$ be the field of displacements of the crystal atoms in cell $x$ ($x\in\Z^d$) from the
equilibrium position.
In the harmonic approximation, the field $u(x)$ is governed by the equations of a type
(see, e.g., \cite{P, LL})
  \beqn\label{1.1'}
\left\{\ba{l}
\ddot u(x,t)=-\sum\limits_{y\in\Z^d}
  V(x-y) u(y,t),\quad x \in\Z^d,\quad t\in\R,\\
u|_{t=0} = u_{0}(x),\quad \dot u|_{t=0} = v_{0}(x).
\ea\right.
\eeqn
Here
 $u(x,t)=(u_1(x,t),\dots,u_n(x,t)), u_0=(u_{01},\dots,u_{0n}),
v_0=(v_{01},\dots,v_{0n})\in\R^n$, $ V(x)$ is the real interaction (or force) matrix,
$ \left(V_{kl}(x)\right)$, $k,l=1,...,n$.
Physically $n=d\times\!$(number of
atoms in the unit cell). Here we take $n$ to be an arbitrary
positive integer.
The dynamics (\ref{1.1'}) is invariant under lattice translations.

Let us denote
$Y(t)=(Y^0(t),Y^1(t))\equiv
(u(\cdot,t),\dot u(\cdot,t))$, $Y_0=(Y^0_0,Y^1_0)\equiv
(u_0(\cdot),v_0(\cdot))$.
Then (\ref{1.1'}) takes the form of an evolution equation
\be\label{CP}
\dot Y(t)={\cal A}(Y(t)),\quad t\in\R;\quad Y(0)=Y_0.
\ee
Formally, this is a linear  Hamiltonian system, since
\be\label{A}
{\cal A}(Y)=J\left( \begin{array}{cc}
{\cal V}& 0\\ 0 & I
\end{array}\right)Y =J\nabla H(Y),\quad
J=\left( \begin{array}{cc}
0 & I\\
- I & 0\end{array}\right).
\ee
Here ${\cal V}$ is a convolution operator with
the matrix kernel $V$, $I$ is unit matrix,
and $H$ is the Hamiltonian functional
\be\label{H}
H(Y):= \frac{1}{2} \langle v,v\rangle
+\frac{1}{2} \langle u,{\cal V}u\rangle, \quad Y=(u,v),
\ee
where the kinetic energy is given by
$(1/2) \langle v,v\rangle=(1/2)\sum\limits_{x\in\Z^d}|v(x)|^2$
and the potential energy by
$(1/2)\langle u,{\cal V}  u\rangle=(1/2)\sum\limits_{x,y\in\Z^d}
\Big(  u(x), V(x-y)u(y)\Big)$,
 $(\cdot,\cdot)$  stands for the real scalar product
 in the Euclidean space $\R^n$ (or in $\R^d$).

We assume that the initial datum $Y_0$
belongs to the phase space ${\cal H}_\alpha$,  $\alpha\in\R$.
 \begin{definition}                 \label{d1.1}
  $ {\cal H}_\alpha$  is the Hilbert space
of pairs $Y\equiv(u(x),v(x))$  of
 $\R^n$-valued functions  of $x\in\Z^d$  endowed  with  the  norm
 \beqn \label{1.5}
 \Vert Y\Vert^2_{\alpha} \equiv
 \sum\limits_{x\in\Z^d}\langle x\rangle^{2\alpha}\Big(\vert u(x)\vert^2
 +\vert v(x)\vert^2\Big) <\infty,
 \quad \langle x\rangle:=\sqrt{1+|x|^2}.
 \eeqn
  \end{definition}

We impose the following conditions {\bf E1}--{\bf E6}  on the matrix $V$.
\begin{itemize}
  \item [{\bf E1}] There exist positive constants $C$ and $\gamma$ such that
$\|V(x)\|\le C e^{-\gamma|x|}$ for $x\in \Z^d$,
$\|V(x)\|$ denoting the matrix norm.
 \end{itemize}
Let $\hat V(\theta)$ be the Fourier transform of $V(x)$, with the convention
$$
 \hat V(\theta)=F_{x\to\theta}[V(x)]\equiv
\sum\limits_{x\in\Z^d}e^{i(x,\theta)}V(x)\,,\quad\theta \in \T^d,
 $$
where   $\T^d$ denotes the $d$-torus $\R^d/(2\pi \Z)^d$.
\begin{itemize}
\item [{\bf E2}]
 $ V$ is real and symmetric, i.e., $V_{lk}(-x)=V_{kl}(x)\in \R$,
$k,l=1,\dots,n$, $x\in \Z^d$.
\end{itemize}
The conditions {\bf E1} and {\bf E2} imply that $\hat V(\theta)$ is a real-analytic
 Hermitian matrix-valued function in $\theta\in \T^d\!$.
 \begin{itemize}
 \item [{\bf E3}]
  The matrix $\hat V(\theta)$ is  non-negative definite for
every $\theta \in \T^d$.
\end{itemize}
Let us define the Hermitian  non-negative definite matrix,
 \be\label{Omega}
 \Omega(\theta)=\big(\hat V(\theta )\big)^{1/2}\ge 0.
 \ee
$\Omega(\theta)$  has the eigenvalues (``dispersion relations'')
$0\leq\omega_1(\theta)<\omega_2(\theta)< \ldots <\omega_s(\theta)$,
$s\leq n$,  and the
corresponding spectral projections $\Pi_\sigma(\theta)$ with
multiplicity $r_\sigma=\tr\Pi_\sigma(\theta)$.
\begin{lemma}\label{lc*} (see \cite[Lemma 2.2]{DKS1}).
Let the conditions {\bf E1} and {\bf E2}
be fulfilled. Then there exists a closed subset ${\cal C}_*\subset \T^d$
of the zero  Lebesgue measure such that  the following assertions hold.
(i) For any point $\Theta\in \T^d\setminus{\cal C}_*$, there
exists a neighborhood ${\cal O}(\Theta)$ such that each band
function $\omega_\sigma(\theta)$ can be chosen as the real-analytic function in
${\cal O}(\Theta)$.\\
(ii) The eigenvalue $\omega_\sigma(\theta)$ has constant multiplicity
in $\T^d\setminus{\cal C}_*$.\\
(iii) The spectral decomposition holds,
 \be\label{spd'}
\Omega(\theta)=\sum_{\sigma=1}^s \omega_\sigma
(\theta)\Pi_\sigma(\theta),\quad \theta\in \T^d\setminus{\cal C}_*,
 \ee
 where $\Pi_\sigma(\theta)$ is the orthogonal projection in
$\R^n$. $\Pi_\sigma$ is a real-analytic function on
$\T^d\setminus{\cal C}_*$.
\end{lemma}

Below we suggest that $\omega_\sigma(\theta)$ denote the local
real-analytic functions from Lemma \ref{lc*} (i).
The next condition on $V$ is the following:
 \begin{itemize}
  \item [{\bf E4}]
 For each $l=1,\dots,d$ and $\sigma=1,\ldots,s$,
$\partial_{\theta_l}\omega_\sigma(\theta)$ does not vanish identically on
$\T^d\setminus{\cal C}_*$.
\end{itemize}

To prove the convergence (\ref{1.8i}), we need a stronger condition {\bf E4'}.
 \begin{itemize}
  \item [{\bf E4'}]
 For each  $\sigma=1,\ldots,s$,
the determinant of the matrix of second partial derivatives of $\omega_\sigma(\theta)$
does not vanish identically on $\T^d\setminus{\cal C}_*$.
\end{itemize}

Write
 \be\label{c0ck}
{\cal C}_0=\{\theta\in \T^d:\det \hat V(\theta)=0\},\,\,
{\cal C}_\sigma=\bigcup_{l=1}^{d}
\{\theta\in \T^d\setminus {\cal C}_*:\,\partial_{\theta_l}\omega_\sigma(\theta)=0\},
\,\,\, \sigma=1,\dots,s.
 \ee
Then the Lebesgue measure of ${\cal C}_\sigma$ vanishes, $\sigma=0,1,...,s$
(see \cite[Lemma 2.3]{DKS1}).
 \begin{itemize}
  \item [{\bf E5}]
    For each $\sigma\ne \sigma'$,
the identities $\omega_\sigma(\theta)
\pm\omega_{\sigma'}(\theta)\equiv\const_\pm$,
$\theta\in \T^d\setminus {\cal C}_*$, do not hold  with $\const_\pm\ne 0$.
\end{itemize}

The condition {\bf E5} can be weakened to the condition~{\bf E5'}, see Remark~\ref{remi-iii} below.
 \begin{itemize}
  \item [{\bf E6}]
 $\Vert \hat V^{-1}(\theta)\Vert\in L^1(\T^d)$.
\end{itemize}
\begin{example}
{\rm
For any $d,n\ge 1$  we  consider {\em the nearest neighbor crystal}  for which
\be\label{dKG}
\langle u,{\cal V}u \rangle=\sum\limits_{l=1}^n\sum\limits_{x\in\Z^d}
\Big(\sum\limits_{i=1}^{d}\kappa_l|u_l(x+e_i)-u_l(x)|^2+
m_l^2|u_l(x)|^2\Big), \,\,\,\kappa_l>0,\,\,\,m_l\ge0,
\ee
where $e_i=(\delta_{i1},\dots, \delta_{id})$.
Then
$$
V_{kl}(x)=0\quad \mbox{for }\,\,k\not=l,\quad V_{ll}(x)=\left\{\ba{ll}
-\kappa_l &\mbox{for }\,\, |x|=1,\\
2d\kappa_l+m^2_l &\mbox{for }\,\, x=0,\\
0 & \mbox{for }\,\,|x|\ge2,
\ea \right. \quad l=1,\dots,n.
$$
Hence,
 the eigenvalues of  $\hat V(\theta)$ are
\be\label{omega}
 \tilde{\omega}_l(\theta)=
\sqrt{\, 2\kappa_l(1-\cos\theta_1)+...+2\kappa_l(1-\cos\theta_d)+m_l^2},\quad l=1,\dots,n.
\ee
These eigenvalues still have to be labelled according to magnitude
and degeneracy as in Lemma~\ref{lc*}.
Clearly the conditions {\bf {E1}}--{\bf {E5}} hold and  ${\cal C}_*=\emptyset$.
If all $m_l>0$, then the set ${\cal C}_0$
 is empty and the condition {\bf E6} is fulfilled.
Otherwise, if $m_l=0$ for some $l$, then ${\cal C}_0=\{0\}$. In this case,
{\bf E6} is equivalent to the condition $\omega_l^{-2}(\theta)\in L^1(\T^d)$
that holds if $d\ge 3$.
Therefore, the  conditions {\bf E1}--{\bf E6}
hold  for (\ref{dKG}) provided either
i) $d\ge 3$, or ii) $d=1,2$ and all $m_l>0$.
}
\end{example}

The following Proposition \ref{p1.1} is proved in
\cite[p.150]{LL}, \cite[p.128]{BPT}.
 \begin{pro}    \label{p1.1}
Let the conditions {\bf E1} and {\bf E2} hold,
and choose some $\alpha\in\R$. Then  for any  $Y_0 \in {\cal H}_\alpha$
 there exists  a unique solution $Y(t)\in C(\R, {\cal H}_\alpha)$
 to the Cauchy problem (\ref{CP});
the operator $U(t):Y_0\mapsto Y(t)$ is continuous in ${\cal H}_\alpha$.
\end{pro}

We assume that $Y_0$ in (\ref{CP}) is a {\em measurable  random function}
and denote by $\mu_0$ a Borel probability measure on ${\cal H}_\alpha$
giving the distribution of  $Y_0$.
Expectation with respect to $\mu_0$ is denoted by $\E$.
We impose the following conditions {\bf S1}--{\bf S3} on the initial measure $\mu_0$.
\begin{itemize}
\item [{\bf S1}]
 $\mu_0$ has  zero expectation value,
$\E(Y_0(x))\equiv \int (Y_0(x))\,\mu_0(dY_0)= 0$, $x\in\Z^d$.

\item [{\bf S2}] The initial correlation functions
$Q^{ij}_0(x,y):= \E\left(Y_0^i(x)\otimes Y_0^j(y) \right)$, $x,y\in\Z^d$,
satisfy the bound
\be\label{Qh}
|Q^{ij}_0(x,y)|\le h(|x-y|),\quad \mbox{where }\,\,r^{d-1} h(r)\in L^1(0,+\infty).
\ee
Here for $a,b,c\in \CC^n$, we denote by $a\otimes b$ the
 linear operator $(a\otimes b)c=a\sum^n_{j=1}b_j c_j$.

\item [{\bf S3}]
Choose some $k\in\{1,\dots,d\}$.
The initial covariance $Q_0(x,y)=\left(Q^{ij}_0(x,y)\right)_{i,j=0,1}$
 depends on difference $x_l-y_l$ for all $l=k+1,\dots,d$, i.e.,
\be\label{3.1}
Q_0(x,y)=q_0(\bar x,\bar y,\tilde x-\tilde y),
\ee
where $x= (x_1,\dots,x_d)\equiv(\bar x,\tilde x)$,
$\bar x=(x_1,\dots,x_k)$, $\tilde x=(x_{k+1},\dots,x_d)$.
Write
\be\label{N}
{\cal N}^k:=\{{\bf n}=(n_1,\dots,n_k),\,\,\,\mbox{where all }\,\,\, n_j\in\{1,2\}\}.
\ee
Suppose that  $\forall\ve>0$ there exists $N(\ve)\in\N$
such that for any
$\bar y\in\Z^k$: $(-1)^{n_j}y_j>N(\ve)$ for each $j=1,\dots,k$,
the following bound holds
\be\label{3a}
\left|q_0(\bar y+\bar z,\bar y,\tilde z)-q_{\bf n}(z)\right|<\ve
\quad\mbox{for any fixed }\,z=(\bar z,\tilde z)\in\Z^d\quad\mbox{and }\,{\bf n}\in{\cal N}^k.
\ee
Here
$q_{\bf n}(z)$, ${\bf n}\in{\cal N}^k$, are  the correlation matrices of some
translation-invariant measures $\mu_{\bf n}$
with zero mean value  in ${\cal H}_\alpha$.
\end{itemize}

In particular, if $k=1$, then  the condition {\bf S3} means that
$Q_0(x,y)=q_0(x_1, y_1,\tilde x-\tilde y)$,
where $x= (x_1,\tilde x)$, $\tilde x=(x_{2},\dots,x_d)$, and
\be\label{1.7'}
q_0(y_1+z_1, y_1,\tilde z)\to\left\{\ba{lll}
q_1(z) &\mbox{ as}& y_1\to-\infty,\\
q_2(z)&\mbox{ as}& y_1\to+\infty,\ea \right.
 \,\,\, z=(z_1,\tilde z)\in\Z^d.
\ee
A measure $\mu$ is called  {\em translation invariant } if
$\mu(T_h B)= \mu(B)$ for  $B\in{\cal B}({\cal H}_{\alpha})$ and
$h\in\Z^d$, where $T_h Y(x)= Y(x-h)$, $x\in\Z^d$,
 ${\cal B}({\cal H}_\alpha)$ stands for the Borel $\sigma$-algebra in ${\cal H}_\alpha$.
Note that the initial measure  $\mu_0$ is not translation-invariant
if $q_{\bf n}\not=q_{{\bf n}'}$ for some ${\bf n}\not={\bf n}'$.
The examples of $\mu_0$ satisfying the conditions {\bf S1}--{\bf S3}
are given in Sec.~\ref{sec3}.

\subsection{Convergence of correlations functions}\label{sec2.3}
\begin{definition}
$\mu_t$ is a Borel probability measure in
${\cal H}_\alpha$ which gives the distribution of $Y(t)$,
$\mu_t(B)=\mu_0(U(-t)B)$, $\forall B\in {\cal B}({\cal H}_\alpha)$, $t\in \R$.
The correlation functions of the  measure $\mu_t$ are  defined by
\be\label{qd}
Q_t^{ij}(x,y)= \E \left(Y^i(x,t)\otimes
 Y^j(y,t)\right),\quad i,j= 0,1,\quad \,x,y\in\Z^d.
\ee
Here $Y^i(x,t)$ are the components of the random solution
$Y(t)=(Y^0(\cdot,t),Y^1(\cdot,t))$.

Denote by   ${\cal Q}_t$  the quadratic form
with the matrix  kernel $(Q^{ij}_t(x,y))_{i,j=0,1}$,
$$
{\cal Q}_t (\Psi, {\Psi})=
\int\left|\langle Y,\Psi\rangle\right|^2\mu_t(dY)=
\sum\limits_{i,j=0,1}~\sum\limits_{x,y\in\Z^d}
\left(Q_t^{ij}(x,y),\Psi^i(x)\otimes\Psi^j(y)\right),\quad t\in\R,
$$
$\Psi=(\Psi^0,\Psi^1)\in{\cal S}:=S\oplus S$,
$S:=S(\Z^d)\otimes \R^n$,
where $S(\Z^d)$ denotes a space of real  quickly decreasing  sequences,
$\langle Y,\Psi \rangle
=\sum\limits_{i=0,1}\sum\limits_{x\in\Z^d}\left( Y^i(x),\Psi^i(x)\right)$.
\end{definition}

Let us introduce the limiting
correlation matrix $Q_\infty(x,y)=(Q^{ij}_\infty(x,y))_{i,j=0}^1$ as follows
\be\label{1.13}
Q_\infty(x,y)=q_\infty(x-y),\quad x,y\in\Z^d.
\ee
Here $q_\infty(x)$ has a form (in the Fourier transform)
\beqn\label{1.15}
 \hat q_{\infty}(\theta)&=& \sum\limits_{\sigma=1}^s
\Pi_\sigma(\theta)({\bf M}^+_{k,\sigma}(\theta)+i\,{\bf M}^-_{k,\sigma}(\theta))\Pi_\sigma(\theta),\quad
\theta\in\T^d\setminus {\cal C}_*,\,
\eeqn
where $\Pi_\sigma(\theta)$ is the spectral projection from Lemma \ref{lc*} (iii),
\beqn\label{2.19}
\ba{lll}
{\bf M}^+_{k,\sigma}(\theta)&=& \ds
\frac{1}{2^{k}}\sum\limits_{{\bf n}\in{\cal N}^k}
L_1^+(\hat q_{\bf n}(\theta))\left[1+S^{{\rm even}}_{k,{\bf n}}(\omega_\sigma(\theta))\right],
\\
{\bf M}^-_{k,\sigma}(\theta)&=&\ds
\frac{1}{2^{k}}\sum\limits_{{\bf n}\in{\cal N}^k}
L_2^-(\hat q_{\bf n}(\theta))\,S^{{\rm odd}}_{k,{\bf n}}(\omega_\sigma(\theta)),
\ea
\eeqn
\beqn\label{Sodd}
 \ba{ll}
S^{{\rm even}}_{k,{\bf n}}(\omega_\sigma)=\!\! \ds
\sum\limits_{{\rm even}\, m\in\{1,\dots, k\}}
\sum\limits_{(p_1,\dots,p_m)\in {\cal P}_m(k)}\ds
\!\!\!\!\!\!\sgn \left( \frac{\partial\omega_\sigma(\theta)}{\partial\theta_{p_1}}\right)\cdot\dots
  \cdot\ds
\sgn \left(\frac{\partial \omega_\sigma(\theta)}{\partial \theta_{p_m}}\right)
(-1)^{n_{p_1}+\dots+n_{p_m}}\\
S^{{\rm odd}}_{k,{\bf n}}(\omega_\sigma)=\!\!
 \ds
\sum\limits_{{\rm odd}\, m\in\{1,\dots, k\}}
\sum\limits_{(p_1,\dots,p_m)\in {\cal P}_m(k)}\ds
\!\!\!\!\!\!\sgn \left( \frac{\partial\omega_\sigma(\theta)}{\partial\theta_{p_1}}\right)\cdot\dots
  \cdot\ds
\sgn \left(\frac{\partial \omega_\sigma(\theta)}{\partial \theta_{p_m}}\right)
(-1)^{n_{p_1}+\dots+n_{p_m}}
\ea
\eeqn
${\cal P}_m(k)$ denotes the collection of all $m$-combinations of the set $\{1,\dots,k\}$
(for instance, ${\cal P}_2(3)=\{(1,2), (2,3), (1,3)\}$),
\beqn\label{1.14}
L_1^\pm(\hat q_{\bf n}(\theta))=\frac12\left(\hat q_{\bf n}(\theta)
\pm C(\theta)\hat q_{\bf n}(\theta)C^*(\theta)\right),\quad
L_2^\pm(\hat q_{\bf n}(\theta))=\frac{1}{2} \left(C(\theta) \hat q_{\bf n}(\theta) \pm
 \hat q_{\bf n}(\theta)C^*(\theta)\right),
 \eeqn
\be\label{C(theta)}
C(\theta)=\left(\ba{cc}
0&\Omega(\theta)^{-1}\\
-\Omega(\theta)&0 \ea\right),\quad
C^*(\theta)=\left(\ba{cc}
0&-\Omega(\theta)\\
\Omega(\theta)^{-1}&0 \ea\right).
 \ee

In particular, if $k=1$, then  formulas (\ref{2.19}) become
\be\label{2.22}
{\bf M}^+_{1,\sigma}(\theta)=\frac12\, L_1^+\left(\hat q_2(\theta)+\hat q_1(\theta)\right),\quad
{\bf M}^-_{1,\sigma}(\theta)=
\frac12\, L_2^-\left(\hat q_2(\theta)-\hat q_1(\theta)\right)
\sgn \left(\frac{\partial\omega_\sigma(\theta)}{\partial{\theta_1}}\right).
\ee
For $d=n=1$,  formulas (\ref{2.22}) were obtained in \cite[p.139]{BPT}.
For any $d,n\ge1$ and $k=1$, these formulas were derived  in \cite{DKM1}.

Note that $\hat  q_\infty\in L^1(\T^d)$ by Lemma~\ref{l4.1}
and the condition~{\bf E6}. Moreover, by (\ref{1.15})--(\ref{C(theta)}),
the matrix $\hat q_\infty(\theta)$ satisfies the ``equilibrium condition'', i.e.,
$\hat q^{11}_\infty(\theta)=\hat V(\theta)\hat q^{00}_\infty(\theta)$,
$\hat q^{10}_\infty(\theta)=-\hat q^{01}_\infty(\theta)$.
Also, $(\hat q^{ii}_\infty(\theta))^*=\hat q^{ii}_\infty(\theta)\!\ge\!0$, $i=0,1$,
$(\hat q^{10}_\infty(\theta))^*=-\hat q^{10}_\infty(\theta)$.

The first result of the paper is the following theorem.
\begin{theorem}\label{tA}
  Let $d,n\ge 1$, $\alpha<-d/2$, and assume that the conditions
{\bf E1}--{\bf E6} and   {\bf S1}--{\bf S3}  hold.  Then
the convergence (\ref{corf}) is true, where $Q_\infty$ is defined in (\ref{1.13})--(\ref{C(theta)}).
\end{theorem}

In Sec.~\ref{sec7}, we study the initial boundary value problem
for harmonic crystals with zero boundary condition
and obtain the results similar to Theorem \ref{tA}, see Theorem \ref{tA+} below.

\begin{remark}\label{remi-iii}
{\rm The condition {\bf E5} on the matrix $V$ could be weakened.
Namely, it suffices to impose the following restriction.
\begin{itemize}
\item [{\bf E5'}]
If for some $\sigma\not=\sigma'$,
 $\omega_\sigma(\theta)+\omega_{\sigma'}(\theta)\equiv \const_+$
with $\const_+\not=0$,
then
$p_{{\bf n},\sigma\sigma'}^{11}(\theta)-\omega_\sigma(\theta)\omega_{\sigma'}(\theta)
p_{{\bf n},\sigma\sigma'}^{00}(\theta)\equiv 0$ and
$\omega_\sigma(\theta)p_{{\bf n},\sigma\sigma'}^{01}(\theta)+
\omega_{\sigma'}(\theta)p_{{\bf n},\sigma\sigma'}^{10}(\theta)\equiv 0$.
If for some $\sigma\not=\sigma'$,
$\omega_\sigma(\theta)-\omega_{\sigma'}(\theta)\equiv \const_-$
with $\const_-\not=0$,
then
$p_{{\bf n},\sigma\sigma'}^{11}(\theta)+\omega_\sigma(\theta)\omega_{\sigma'}(\theta)
p_{{\bf n},\sigma\sigma'}^{00}(\theta)\equiv 0$ and
$\omega_\sigma(\theta)p_{{\bf n},\sigma\sigma'}^{01}(\theta)-
\omega_{\sigma'}(\theta)p_{{\bf n},\sigma\sigma'}^{10}(\theta)\equiv 0$.
Here
\be\label{pij}
p_{{\bf n},\sigma\sigma'}^{ij}(\theta):=
\Pi_\sigma(\theta)\hat q_{\bf n}^{ij}(\theta)\Pi_{\sigma'}(\theta),\,\,\,
\theta\in \T^d,\,\,\, \sigma,\sigma'=1,\dots,s,\,\,\, i,j=0,1,\,\,\, {\bf n}\in{\cal N}^k.
\ee
\end{itemize}
This condition  holds, for instance, for the canonical Gibbs measures $\mu_{\bf n}$
considered in Sec.~\ref{sec4.2}.
}
\end{remark}
\begin{examples}
{\rm
We rewrite the formulas for $q_\infty$ in some particular cases.

(i)\,
In the case when  the initial covariance is translation invariant, i.e.,
$Q_0(x,y)=q_0(x-y)$, the matrix $\hat q_\infty$ is of a form
\be\label{2.2}
 \hat q_{\infty}(\theta)= \sum\limits_{\sigma=1}^s
\Pi_\sigma(\theta)L^+_{1}(\hat q_0(\theta))\Pi_\sigma(\theta),\quad
\theta\in\T^d\setminus {\cal C}_*.
\ee

(ii)\,
Let  the initial covariance $Q_0$ satisfy a stronger condition than (\ref{3a}).
Namely, assume that $Q_0$ has a form (\ref{3.1}) and for any $z=(\bar z,\tilde z)\in\Z^d$,
$
\lim_{|\bar y|\to\infty}q_0(\bar y+\bar z,\bar y, \tilde z)=q_*(z).
$
Then the condition (\ref{3a}) is fulfilled with $q_{\bf n}(z)=q_*(z)$
for any ${\bf n}\in{\cal N}^k$.
In this case, Theorem~\ref{tA} holds, and
 $\hat q_\infty$ is of a form (\ref{2.2}) with $\hat q_*$ instead of $\hat q_0$.
Therefore, Theorem~\ref{tA} generalizes the result of \cite[Proposition 3.2]{DKS1},
where the convergence (\ref{corf}) was proved in the case when
$Q_0(x,y)=q_0(x-y)$.
}\end{examples}

\subsection{Weak convergence of measures}\label{sec2.4}

To prove the  convergence  (\ref{1.8i}) of the measures $\mu_t$,
we impose a stronger condition {\bf S4} on $\mu_0$
than the bound (\ref{Qh}).
To formulate this condition, let us denote by $\sigma ({\cal A})$,
${\cal A}\subset \Z^d$, the $\sigma $-algebra in
${\cal H}_{\alpha}$ generated by $Y_0(x)$ with $x\in{\cal A}$.
Define the Ibragimov mixing coefficient
of a probability  measure  $\mu_0$ on ${\cal H}_\alpha$
by the rule (cf \cite[Definition 17.2.2]{IL})
$$
\varphi(r)\equiv
\sup_{\ba{c}{\cal A},{\cal B}\subset \Z^d:\\{\rm dist}({\cal A},{\cal B})\ge r\ea}
 \sup_{ \ba{c} A\in\sigma({\cal A}),B\in\sigma({\cal B})\\ \mu_0(B)>0\ea}
\fr{| \mu_0(A\cap B) - \mu_0(A)\mu_0(B)|}{ \mu_0(B)}.
$$
\begin{definition}\label{mix-cond}
 The measure $\mu_0$  satisfies a strong uniform
Ibragimov mixing condition if $\varphi(r)\to 0$ as $r\to\infty$.
\end{definition}
\begin{itemize}
  \item [{\bf S4}]
The initial mean ``energy'' density  is uniformly bounded:
\be\label{med}
 \E [\vert u_0(x)\vert^2+\vert v_0(x)\vert^2]
={\rm tr}\,Q_0^{00}(x,x)+{\rm tr}\,Q_0^{11}(x,x)
\le e_0<\infty,\quad x\in\Z^d.
\ee
Moreover,
$\mu_0$ satisfies the  strong uniform  Ibragimov mixing condition and
$$
 \int_0^\infty r^{d-1}\varphi^{1/2}(r)\,dr<\infty.
$$\end{itemize}
\begin{remark}
{\rm By \cite[Lemma 17.2.3]{IL}, the conditions {\bf S1} and {\bf S4}  imply
the bound (\ref{Qh}) with $h(r)=Ce_0 \varphi^{1/2}(r)$,
where $e_0$ is a constant from the bound (\ref{med}).
}\end{remark}

For a probability  measure $\mu$ on  ${\cal H}_\alpha$
we denote by $\hat\mu$ the characteristic functional (Fourier transform),
$\hat\mu(\Psi)\equiv\int\exp(i\langle Y,\Psi\rangle)\,\mu(dY)$, $\Psi\in{\cal S}$.
A measure $\mu$ is called {\em Gaussian} (of zero mean) if
its characteristic functional has the form
$\hat{\mu}(\Psi)=  \exp\{-{\cal Q}(\Psi,\Psi)/2\}$,
where ${\cal Q}$ is a real nonnegative quadratic form in ${\cal S}$.
\begin{theorem}\label{tB}
  Let $d,n\ge 1$, $\alpha<-d/2$, and assume that the conditions
{\bf E1}--{\bf E3}, {\bf E4'}, {\bf E5'}, {\bf E6}, {\bf S1}, {\bf S3}, and {\bf S4}  be fulfilled.
Then the following assertions hold.\\
(i) The measures  $\mu_t$ weakly converge in  the Hilbert space ${\cal H}_\alpha$,
\be\label{1.8}
\mu_t\to \mu_\infty \quad as\quad t\to \infty.
\ee
The limit measure $ \mu_\infty $ is a Gaussian translation-invariant
measure on ${\cal H}_\alpha$.
The  characteristic functional of $ \mu_{\infty}$ is of a form
$\ds\hat {\mu}_\infty(\Psi)=
\exp\{-\fr{1}{2}{\cal Q}_\infty (\Psi,\,\Psi)\}$, $\Psi\in{\cal S}$,
where ${\cal Q}_\infty$ is the quadratic form with the matrix kernel $Q_\infty(x,y)$
defined in (\ref{1.13}).\\
(ii) The measure $\mu_\infty$ is time stationary, i.e.,
$[U(t)]^*\mu_\infty=\mu_\infty$, $t\in\R$.\\
(iii) The flow $U(t)$ is mixing with respect to  the measure $\mu_\infty$, i.e.,
for any $f,g\in L^2({\cal H}_{\alpha},\mu_\infty)$,
$$
\lim_{t\to\infty}
\int f(U(t)Y)g(Y)\,\mu_{\infty}(dY)=\int f(Y)\,\mu_{\infty}(dY)
\int g(Y)\,\mu_{\infty}(dY)
$$
In particular, the flow $U(t)$ is ergodic with respect to  the measure $\mu_\infty$.
\end{theorem}

For harmonic crystals in the half-space, the convergence (\ref{1.8}) also holds.
For details, see Sec.~\ref{sec7}.
The assertion  (i) of Theorem~\ref{tB} follow
from Propositions  \ref{l2.1} and \ref{l2.2}.
 \begin{pro}\label{l2.1}
Let the  conditions {\bf E1}--{\bf E3}, {\bf E6}, {\bf S1} and {\bf S2} hold.
Then the measures family  $\{\mu_t,\,t\in \R\}$
 is weakly compact in   ${\cal H}_\alpha$ with any  $\alpha<-d/2$,
and the following bounds hold
 \beqn \label{20.1}
\sup\limits_{t\in\R}
\E\Vert U(t)Y_0\Vert^2_\alpha<\infty.
\eeqn
 \end{pro}
 \begin{pro}\label{l2.2}
Let the conditions
 {\bf E1}--{\bf E3}, {\bf E4'}, {\bf E5'}, {\bf E6}, {\bf S1}, {\bf S3}, and {\bf S4}  hold.
Then for every $\Psi\in {\cal S}$, the characteristic functionals of $\mu_t$
  converge to a Gaussian one,
$$
 \hat\mu_t(\Psi):=\int e^{i\langle Y,\Psi\rangle}\mu_t(dY)
\rightarrow \exp\{-\fr{1}{2}{\cal Q}_\infty (\Psi ,\Psi)\},\quad t\to\infty.
$$
\end{pro}

Proposition~\ref{l2.1} (Proposition~\ref{l2.2}) provides the existence
(resp. the uniqueness) of the limit measure $\mu_\infty$.
Proposition  \ref{l2.1} is proved in Sec.~\ref{sec5}.
Proposition \ref{l2.2} can be proved using the technique from \cite{DKM1}.
The assertion~(ii) of Theorem~\ref{tB}  follows from (\ref{1.8})
since the group $U(t)$ is continuous in ${\cal H}_\alpha$ by Proposition~\ref{p1.1}.
The ergodicity and mixing of the limit measures
$\mu_\infty$ follow by the same arguments as in \cite{DKS1}.
\begin{lemma}\label{tB0}
Let  the conditions
{\bf E1}--{\bf E4},  {\bf E5'}, and {\bf E6} hold.
Assume that the initial measure $\mu_0$ is Gaussian and satisfies the
conditions {\bf S1}--{\bf S3}.
Then all assertions of Theorem~\ref{tB} remain valid.
\end{lemma}

This lemma follows from Theorem~\ref{tA} and Proposition~\ref{l2.1}.

\setcounter{equation}{0}
\section{Examples of initial measures} \label{sec3}

Now we construct
Gaussian initial measures $\mu_0$ satisfying the conditions {\bf S1}--{\bf S3}.
For $k=1$ (see the condition~{\bf S3}), the example of $\mu_0$ is given in \cite{DKM1}.
For any  $k\ge1$,  the measure  $\mu_0$ can be constructed by a following way.
At first, for simplicity, we assume that $u_0,v_0\in\R^1$ and  define  the
 correlation functions $q_{\bf n}^{ij}(x-y)$, ${\bf n}\in{\cal N}^k$,
  which are zero for  $i\not= j$, while for $i=0,1$,
\be\label{S04}
\hat q_{\bf n}^{ii}(\theta):=F_{z\to\theta}
[ q_{\bf n}^{ii}(z)]\in L^1(\T^d),\quad \hat q_{\bf n}^{ii}(\theta) \ge 0.
\ee
Then, by the Minlos theorem \cite{CFS}, there exist  Borel Gaussian measures
$\mu_{\bf n}$ on  ${\cal H}_\alpha$,
$\alpha<-d/2$, with the correlation functions $q^{ij}_{\bf n}(x-y)$,
because
$$
\int\Vert Y\Vert^2_\alpha\,\mu_{\bf n}(dY)
=\sum\limits_{x\in\Z^d}\langle x\rangle^{2\alpha}
\tr\left(q^{00}_{\bf n}(0)+ q_{\bf n}^{11}(0)\right) =C(\alpha,d)\int_{\T^d}
\tr\left(\hat q^{00}_{\bf n}(\theta)+\hat q_{\bf n}^{11}(\theta)\right)d\theta<\infty.
$$
Further, we take the functions $\ov\zeta_{\bf n}\in C(\Z^k)$ such that
$$
\ov\zeta_{\bf n}(\bar x)=\zeta_{n_1}( x_1)\cdot\dots\cdot\zeta_{n_k}( x_k),
\quad \bar x=(x_1,\dots,x_k),\quad {\bf n}=(n_1,\dots,n_k),\quad { n}_j\in\{1,2\},
$$
where the sequences $\zeta_1(x)$ and $\zeta_2(x)$, $x\in\Z$, are defined by the rule
\be\label{zeta}
\zeta_{1}(x)= \left\{ \ba{lll}
1&\mbox{for }~ x<\,-a,\\
0&\mbox{for }~ x>a,\ea\right.
\quad \zeta_{2}(x)= \left\{ \ba{lll}
1&\mbox{for }~ x>\,a,\\
0&\mbox{for }~  x<-a,\ea\right.\quad \mbox{with some }\,\, a>0.
\ee
Finally, define a  Borel probability measure
 $\mu_0$ as a distribution of the random function
\be\label{rf-k}
Y_0(x)= \sum_{{\bf n}\in{\cal N}^k}\ov\zeta_{{\bf n}}(\bar x)Y_{\bf n}(x),
\quad x=(\bar x,\tilde x)\in \Z^d, \quad \bar x=(x_1,\dots, x_k),
\quad \tilde x=(x_{k+1},\dots,x_d),
\ee
where $Y_{\bf n}(x)$ are  Gaussian independent functions in ${\cal H}_\alpha$ with distributions $\mu_{\bf n}$.
Then correlation matrix of  $\mu_0$ is of a form
\be\label{q0-k}
Q_0(x,y)=\sum_{{\bf n}\in{\cal N}^k}\ov\zeta_{{\bf n}}(\bar x)\ov\zeta_{{\bf n}}(\bar y)q_{\bf n}(x-y),
\ee
where $x= (\bar x,\tilde x)$, $y= (\bar y, \tilde y)\in \Z^d$,
and $q_{\bf n}(x-y)$ are the correlation matrices  of the measures $\mu_{\bf n}$.
Hence, $Q_0(x,y)=q_0(\bar x,\bar y, \tilde x-\tilde y)$,  and
for every $ z=(\bar z,\tilde z)\in\Z^d$,
$$
q_0(\bar y+\bar z,\bar y,\tilde z)=q_{\bf n}(z)\,\,\,\,\mbox{ if}\,\,\, (-1)^{n_j}y_j>a+|z_j|,
\quad \forall j=1,\dots,k,\quad {\bf n}=(n_1,\dots,n_k).
$$
Therefore, the measure $\mu_0$ satisfies the conditions {\bf S1} and {\bf S3}.
If
\be\label{S5'}
|q_{\bf n}^{ii}(z)|\le h(|z|),\quad \mbox{where }\,\,\,r^{d-1}h(r)\in L^1(0,+\infty),
\ee
then $\mu_0$ satisfies {\bf S2} by (\ref{q0-k}).
Now we give  examples of $q_{\bf n}^{ii}$ satisfying (\ref{S04}) and (\ref{S5'}).
\begin{example}
{\rm
Put $q_{\bf n}^{ii}(z)=f(z_1)f(z_2)\cdot\dots\cdot f(z_d)$
and construct sequences $f(z)$, $z\in\Z$, such that
conditions (\ref{S04}) and (\ref{S5'}) hold.

(i) Let $f(z)=N_0-|z|$ for $|z|\le N_0$ and $f(z)=0$ for $|z|> N_0$  with
 some $N_0>0$. Then
 $q_{\bf n}^{ii}(z)=0$ for $|z|\geq r_0\equiv N_0\sqrt d$,  and
$\hat f(\theta)=(1-\cos N_0\theta)/(1-\cos\theta)$,
$\theta\in \T^1$. Hence,  (\ref{S04}) and (\ref{S5'}) are fulfilled.
Furthermore, the condition {\bf S4}
also follows with  $\varphi(r)=0$ for $r\geq r_0$.
This example of the sequence $f$ can be  generalized as follows.

 Let $f$ be an even nonnegative sequence  such that
 $f\in\ell^1$ and $\Delta_L f(z)\ge0$ for any $z\ge1$. Then
$\hat f(\theta)\ge0$ by \cite[Theorems 4.1 and 2.7]{Kat}, and (\ref{S04}) follows.
If, in addition,  $|f(z)|\le C(1+|z|)^{-N}$ with $N>d$, then
 (\ref{S5'}) holds.
\smallskip

(ii)\,
Let  $f(z)=(a+b|z|)\gamma^{|z|}$, $z\in\Z$,
 with $\gamma\in(0,1)$, $a>0$ and $b\ge0$.
Hence, (\ref{S5'}) is fulfilled with $h(r)=C(1+r)^{-N}$ ($N>d$).
 If $a\ge 2b\gamma/(1-\gamma)$, then $\Delta_L f(z)\ge0$ for any $z\ge1$
 and $\hat f(\theta)\ge0$ (see the case (i)).
If $2b\gamma/(1-\gamma^2)\le a<2b\gamma/(1-\gamma)$, then $\Delta_L f(1)<0$.
However, in this case, (\ref{S04}) also holds because
 $$
  \ba{lll}
  \hat f(\theta)
 &=&\ds\frac{a(1-\gamma^2)}{1-2\gamma\cos\theta+\gamma^2}
 +\frac{2b\gamma\left((1+\gamma^2)\cos\theta-2\gamma\right)}{\left(1-2\gamma\cos\theta+\gamma^2\right)^2}
 \\
 &=&\ds
 \frac{[a(1-\gamma^2)+2b\gamma\cos\theta)](1-\gamma)^2+[a(1-\gamma^2)-2b\gamma](1-\cos\theta)2\gamma}
 {\left(1-2\gamma\cos\theta+\gamma^2\right)^2}\ge0.
 \ea
$$
}\end{example}
\begin{example}\label{ex3}
{\rm
Let $u_0,v_0\in\R^n$ with any $n\ge1$, and
$q^{00}_{\bf n}(z)=T_{\bf n}F^{-1}_{\theta\to z}[\hat V^{-1}(\theta)]$,
 $q^{11}_{\bf n}(z)=T_{\bf n}I$, $z\in\Z^d$, with some constants $T_{\bf n}>0$.
Assume, in addition, that  ${\cal C}_0=\emptyset$ (see (\ref{c0ck})),
i.e.
\be\label{E6'}
\det \hat V(\theta)\not=0,\quad\forall\theta\in \T^d.
\ee
Hence,
\be\label{ft}
|q_{\bf n}^{00}(z)|=T_{\bf n}\left|F^{-1}_{\theta\to z}
[\hat V^{-1}(\theta)]\right|\sim (1+|z|)^{-N},\quad \forall N\in\N,
\ee
and the conditions (\ref{S04})  and (\ref{S5'}) are fulfilled
with $h(r)=(1+r)^{-N}$, where $N> d$.
}
\end{example}
\begin{remark}
{\rm Suppose that the initial covariance has a particular form:
\be\label{3.8}
Q_0(x,y)=T(\bar x+\bar y)r(x-y)\,\,\,\, \mbox{or }\,\,\,Q_0(x,y)=\sqrt {T(\bar x) T(\bar y)}r(x-y),
\ee
where $T(\bar x)$ is a bounded nonnegative sequence on $\Z^k$,
$r(x)=(r^{ij}(x))$ is a correlation matrix of some
translation-invariant measure in ${\cal H}_\alpha$
with zero mean value,
$|r^{ij}(x)|\le h(|x|)$, where $r^{d-1} h(r)\in L^1(0,+\infty)$.
Then, the condition~{\bf S2} is fulfilled.
For every ${\bf n}=(n_1,\dots,n_k)\in{\cal N}^k$,
we assume that $\forall\ve>0$ $\exists N(\ve)\in\N$
such that for any $\bar x\in\Z^k$:
$(-1)^{n_j}x_j>N(\ve)$ with any $j=1,\dots,k$,   $|T(\bar x)-T_{\bf n}|<\ve$.
Hence, the condition~{\bf S3} is fulfilled with $q_{{\bf n}}(x):=T_{\bf n}\,r(x)$.
}\end{remark}

\setcounter{equation}{0}
\section{Energy current} \label{sec4}
\subsection{Non-equilibrium states}

At first, we derive  the expression for the energy current density of
the finite energy solutions $u(x,t)$ (see (\ref{H})).
For the half-space $\Omega_l:=\{x\in\Z^d:\,x_l\ge 0\}$,
we define the energy in the region $\Omega_l$ (cf (\ref{H})) as
$$
{\cal E}_l(t):=\frac12\sum\limits_{x\in\Omega_l}
\Big\{|\dot u(x,t)|^2+\sum\limits_{y\in\Z^d}
\Big(u(x,t),V(x-y)u(y,t)\Big)\Big\},\quad l=1,\dots,d.
$$
Introduce new variables: $x=x'+me_l$, $y=y'+pe_l$,
where $x',y'\in\Z^d$ with $x'_l=y'_l=0$,
$e_l=(\delta_{l1},\dots,\delta_{ld})$, $l=1,\dots,d$.
Using Eqn~(\ref{1.1'}),  we obtain
$\dot {\cal E}_l(t)=\sum\limits_{x'}J^l(x',t)$.
Here $J^l(x',t)$ stands for the  energy current density
in the direction $e_l$:
\beqn
 J^l(x',t):&=&\fr12\sum\limits_{y'}
\Big\{\sum\limits_{m\le-1,\,p\ge 0}
\Big(\dot u(x'+me_l,t),V(x'+me_l-y'-pe_l)u(y'+pe_l,t)\Big)
\nonumber\\
&&-\sum\limits_{m\ge0,\,p\le -1}
\Big(\dot u(x'+me_l,t),V(x'+me_l-y'-pe_l)u(y'+pe_l,t)\Big)\Big\},
\nonumber
\eeqn
where $x',y'\in\Z^d$ with $x'_l=y'_l=0$.
Further, let $u(x,t)$ be the random solution of Eqn (\ref{1.1'}) with
the initial measure $\mu_0$ satisfying {\bf S1}--{\bf S3}.
The  convergence   (\ref{corf}) yields
\beqn
\E \left(J^l(x',t)\right)\to   J^l_\infty
&:=&\fr12 \sum\limits_{y'}\Big(\sum\limits_{m\le-1,\,p\ge 0}
\tr\Big[q^{10}_\infty (x'-y'+(m\!-\!p)e_l) V^T(x'-y'+(m\!-\!p)e_l)\Big]
\nonumber\\
&&-\sum\limits_{m\ge0,\,p\le -1}
\tr\Big[q^{10}_\infty (x'-y'+(m\!-\!p)e_l) V^T(x'-y'+(m\!-\!p)e_l)\Big]
\Big)\nonumber\\
&=&-\fr12 \sum\limits_{z\in\Z^d} z_l
\tr\Big[q^{10}_\infty (z) V^T(z)\Big]\quad \mbox{as }\,\,t\to\infty.
\nonumber
\eeqn
 Applying Fourier transform and the equality
 $\hat V^*(\theta)=\hat V(\theta)$, we obtain
$$
  J^l_\infty = -\fr{(2\pi)^{-d}}{2}
\int_{\T^d} i\,\tr\left[\hat q^{10}_\infty(\theta) \pa_{\theta_l}
\hat V(\theta)\right]\,d\theta, \quad l=1,\dots,d.
$$
Since $\Pi_\sigma(\theta)$ are orthogonal projections,
then $\Pi_\sigma(\theta)\left(\partial_{\theta_l}\Pi_{\sigma'}(\theta)\right)\Pi_\sigma(\theta)=0$
for any $\sigma,\sigma'=1,\dots,s$ and $l=1,\dots,d$.
Hence, applying the formula (\ref{1.15}) and the decomposition of $\hat V(\theta)$,
$\hat V(\theta)=\sum_{\sigma=1}^s\Pi_\sigma(\theta)\omega_\sigma^2(\theta)$,
we obtain that
$\tr\left[\hat q^{10}_\infty(\theta)
\sum_{\sigma=1}^s\omega^2_\sigma(\theta)\pa_{\theta_l}\Pi_\sigma(\theta)\right]=0$
and
\beqn\label{jinfty2}
 J^l_\infty &=& -i(2\pi)^{-d}
 \sum_{\sigma=1}^s
\int_{\T^d} \tr\left[\Pi_{\sigma}(\theta)\left({\bf M}^+_{k,\sigma}(\theta)
+i{\bf M}^-_{k,\sigma}(\theta)\right)^{10}\Pi_{\sigma}(\theta)\right]
\omega_\sigma(\theta)\pa_{\theta_l}\omega_\sigma(\theta)\,d\theta
\nonumber\\
&=&-(2\pi)^{-d}\frac{1}{2^k} \sum_{\sigma=1}^s\sum\limits_{{\bf n}\in{\cal N}^k}
\Big\{
\frac12\int_{\T^d} \tr\left[
\omega^2_{\sigma}(\theta)p^{00}_{\bf n,\sigma\sigma}(\theta)+p^{11}_{\bf n,\sigma\sigma}(\theta)
\right]
S^{{\rm odd}}_{k,{\bf n}}(\omega_\sigma)\pa_{\theta_l}\omega_\sigma(\theta)\,d\theta
\nonumber\\
&&+\int_{\T^d} {\rm Im}\left(\tr p^{01}_{\bf n,\sigma\sigma}(\theta)\right)
\left(1+S^{{\rm even}}_{k,{\bf n}}(\omega_\sigma)\right)\omega_\sigma(\theta)\pa_{\theta_l}\omega_\sigma(\theta)\,d\theta
\Big\}, \quad l=1,\dots,d,
\eeqn
where $p^{ij}_{\bf n,\sigma\sigma'}$ are introduced in (\ref{pij}).
Here we use the equality $\overline{\tr p^{ij}_{\bf n,\sigma\sigma}(\theta)}=\tr p^{ji}_{\bf n,\sigma\sigma}(\theta)$.
\begin{remark}
{\rm
Let, for simplicity, all functions $\tr[ p^{ij}_{\bf n,\sigma\sigma}(\theta)]$
and $\omega_\sigma(\theta)$ be even for every variable $\theta_1,\dots,\theta_d$.
Then,  $J^l_\infty=0$ for $l>k$, and $J^l_\infty=C_1^l-C_2^l$ for $l=1,\dots,k$,
where
$$
C^l_j:=(2\pi)^{-d}\frac{1}{2^k} \sum_{\sigma=1}^s{\sum_{\bf n}}'\,
\frac12\int_{\T^d} \tr\left[
\omega^2_{\sigma}(\theta)p^{00}_{\bf n,\sigma\sigma}(\theta)+p^{11}_{\bf n,\sigma\sigma}(\theta)
\right]\Big|_{n_l=j}\left|\pa_{\theta_l}\omega_\sigma(\theta)\right|\,d\theta,\quad j=1,2.
$$
Here the summation $\sum'_{\bf n}$ is taken over $n_1,\dots,n_{l-1},n_{l+1},\dots,n_k\in\{1,2\}$.
In particular, for the initial correlation matrix $Q_0$ of the form (\ref{3.8}),
 $p^{ij}_{\bf n,\sigma\sigma}(\theta)=T_{\bf n}p^{ij}_{\sigma\sigma}(\theta)$,
where, by definition, $p^{ij}_{\sigma\sigma}(\theta):=\Pi_\sigma(\theta)\hat r^{ij}(\theta)\Pi_\sigma(\theta)$.
In this case, ${\bf J}_\infty$ is of the form (\ref{Jl}) with
 $$
 c_l= \sum_{\sigma=1}^s(2\pi)^{-d}
 \frac12\int_{\T^d} \tr\left[
\omega^2_{\sigma}(\theta)p^{00}_{\sigma\sigma}(\theta)+p^{11}_{\sigma\sigma}(\theta)
\right]\left|\pa_{\theta_l}\omega_\sigma(\theta)\right|\,d\theta.
$$
We see that one can choose numbers $T_{\bf n}$ such that $J^l_\infty\not=0$ for some $l=1,\dots,k$.
}\end{remark}

Below we  simplify the formula (\ref{jinfty2}) in the case when $\mu_{\bf n}$ are  Gibbs measures
corresponding to  positive temperatures $T_{\bf n}$.
Furthermore, under the additional symmetry conditions
on the interaction matrix $V$, we derive the formula (\ref{Jl}) for ${\bf J}_\infty$.
Thus, there exist stationary non-equilibrium  states (in fact, Gaussian measures $\mu_\infty$)
in which there is a non-zero
constant energy current passing through the points of the crystal.

\subsection{Energy current for the Gibbs measures} \label{sec4.2}

Formally, Gibbs measures $g_\beta$ are
$$
g_{\beta}(dY)= \frac{1}{Z}\,
\ds e^{-\ds  \beta H(Y)}\prod_{x\in\Z^d}dY(x),
$$
where $H(Y)$ is defined in (\ref{H}), $Z$ is normalization factor, $\beta=T^{-1}$,
$T>0$ is a  corresponding absolute temperature.
We introduce the Gibbs measures $g_{\beta}(dY)$ as the
Gaussian measures in ${\cal H}_\alpha$, $\alpha<-d/2$,
with zero mean and with the correlation matrices
defined by their Fourier transform,
\be\label{4}
\hat q_{\beta}^{00}(\theta)= T\hat V^{-1}(\theta),\quad
\hat q_{\beta}^{11}(\theta)= TI,\quad
\hat q_{\beta}^{01}(\theta)= \hat q_{\beta}^{10}(\theta)= 0,
\ee
where $I$ denotes  unit matrix in $\R^n\times\R^n$.
By the Minlos theorem \cite{CFS},
the  Borel probability measures $g_{\beta}$
exist in the spaces ${\cal H}_\alpha$. Indeed,
$$
\int \Vert Y\Vert^2_{\alpha}\, g_{\beta}(dY)
=\sum\limits_{x\in\Z^d}\langle x\rangle^{2\alpha}\,
\tr [q^{00}_{\beta}(0)+q^{11}_{\beta}(0)]<\infty,
 $$
since $\alpha<-d/2$ and
$$
\tr [q^{00}_{\beta}(0)+q^{11}_{\beta}(0)]=(2\pi)^{-d}\int_{\T^d}\tr
[\hat q^{00}_{\beta}(\theta)+\hat q^{11}_{\beta}(\theta)]\,d\theta=
T(2\pi)^{-d}\int_{\T^d}\tr \hat V^{-1}(\theta)\,d\theta+Tn<\infty.
$$
The last bound is obvious if ${\cal C}_0=\emptyset$
and  follows from the condition~{\bf E6} if ${\cal C}_0\not=\emptyset$.
\smallskip

 Let $\mu_0$  be a Borel probability measure in ${\cal H}_{\alpha}$
 giving the distribution of the random function $Y_0$
   constructed in Sec.~\ref{sec3} (see formula (\ref{rf-k})) with Gibbs measures
   $\mu_{\bf n}\equiv g_{\beta_{\bf n}}$
($\beta_{\bf n}=1/T_{\bf n}$, $T_{\bf n}>0$)
which have  correlation matrices $q_{\bf n}(x)\equiv q_{\beta_{\bf n}}(x)$,
where the matrix $q_{\beta}=(q^{ij}_\beta)_{i,j=0,1}$ is defined by (\ref{4}).
We impose, in addition, the condition~(\ref{E6'}).
Then, the conditions~{\bf S1}--{\bf S3} hold (see Example~\ref{ex3}).
We check that in the case of the Gibbs measures $\mu_{\bf n}\equiv g_{\beta_{\bf n}}$,
the condition~{\bf E5'}  is fulfilled (see Remark~\ref{remi-iii}).
Indeed, by (\ref{4}) we have
\beqn\label{4.8}
\left.\ba{lll}
p_{{\bf n},\sigma\sigma'}^{00}(\theta)&=&
\Pi_\sigma(\theta)\hat q^{00}_{\bf n}(\theta)\Pi_{\sigma'}(\theta)=
T_{\bf n}\omega_\sigma^{-2}(\theta)\Pi_\sigma(\theta)\delta_{\sigma\sigma'}\\
p_{{\bf n},\sigma\sigma'}^{11}(\theta)&=&
\Pi_\sigma(\theta)\hat q^{11}_{\bf n}(\theta)\Pi_{\sigma'}(\theta)
=T_{\bf n}\,\Pi_\sigma(\theta)\delta_{\sigma\sigma'}\ea \right|
\quad \sigma,\sigma'=1,\dots,s,
\eeqn
and $p_{{\bf n},\sigma\sigma'}^{ij}(\theta)=0$ for
 $i\not=j$.
Therefore, the convergence (\ref{1.8}) holds by Lemma~\ref{tB0}.
\smallskip

Now  we rewrite  the limit covariance
$\hat q_\infty(\theta)$ and  the limit mean energy current ${\bf J}_\infty$ 
in the case when  $\mu_{\bf n}=g_{\beta_{\bf n}}$  are Gibbs measures.
Applying (\ref{1.15}), (\ref{2.19}) and (\ref{1.14}) we obtain
\beqn\nonumber
\ba{lll}\hat q_{\infty}^{11}(\theta)&\!=\!&
\ds\hat V(\theta)\hat q_{\infty}^{00}(\theta)
=\sum\limits_{\sigma=1}^s\Pi_\sigma(\theta)
\,\frac{1}{2^{k}}\sum\limits_{{\bf n}\in{\cal N}^k}
T_{\bf n}\left[1+S^{{\rm even}}_{k,{\bf n}}(\omega_\sigma(\theta))\right],\\
\hat q_{\infty}^{10}(\theta)
&\!=\!&\ds
-\hat q_{\infty}^{01}(\theta)
=-i\,\sum\limits_{\sigma=1}^s \Pi_\sigma(\theta)\omega_\sigma^{-1}(\theta)
\,\frac{1}{2^{k}}\sum\limits_{{\bf n}\in{\cal N}^k}
T_{\bf n}S^{{\rm odd}}_{k,{\bf n}}(\omega_\sigma(\theta)),
\ea
\eeqn
where the functions $S^{{\rm even}}_{k,{\bf n}}(\omega_\sigma)$ and
$S^{{\rm odd}}_{k,{\bf n}}(\omega_\sigma)$
are defined in (\ref{Sodd}).
Substituting $p_{{\bf n},\sigma\sigma}^{ij}(\theta)$ from (\ref{4.8})
 in the r.h.s. of (\ref{jinfty2}), we obtain
\beqn\label{mecd}
 J^l_\infty \!&=&\!-\fr{1}{(2\pi)^{d}}\,\frac{1}{2^{k}}
\sum\limits_{\sigma=1}^{s} \sum\limits_{{\bf n}\in{\cal N}^k} \int_{\T^d}r_\sigma\,
T_{\bf n}\,S^{{\rm odd}}_{k,{\bf n}}(\omega_\sigma(\theta))\,
\pa_{\theta_l}\omega_{\sigma}(\theta)\,d\theta
\nonumber\\
\!&=&\!
- \frac{1}{2^{k}}\sum\limits_{{\bf n}\in{\cal N}^k}
T_{\bf n}\Big(
\sum\limits_{{\rm odd}\, m\in\{1,\dots, k\}}
\sum\limits_{(p_1,\dots,p_m)\in {\cal P}_m(k)}c^l_{p_1\dots p_m}\,
(-1)^{n_{p_1}+\dots+n_{p_m}}\Big),\,\,\,l=1,\dots,d,\,\,\,
\eeqn
where $r_\sigma=\tr[\Pi_{\sigma}(\theta)]$
is multiplicity of the eigenvalue $\omega_\sigma$ (see Lemma \ref{lc*}),
 the numbers $c^l_{p_1\dots p_m}$ are defined as follows
\be\label{4.11}
c^l_{p_1\dots p_m}:=
\fr{1}{(2\pi)^{d}}\sum\limits_{\sigma=1}^{s}\int_{\T^d}r_\sigma \,
{\rm sign} \left( \frac{\partial\omega_\sigma(\theta)}{\partial\theta_{p_1}}\right)\cdot\dots
 \cdot
{\rm sign} \left(\frac{\partial \omega_\sigma(\theta)}{\partial \theta_{p_m}}\right)
\frac{\pa\omega_{\sigma}(\theta)}{\pa\theta_l}\,d\theta.
\ee

Under the additional symmetry conditions ({\bf SC}) on the interaction matrix $V$, the formulas
(\ref{mecd}) and (\ref{4.11})  can be simplified.
\begin{itemize}
\item [{\bf SC}] Suppose  that  one of the following conditions on $\omega_\sigma$, $\sigma=1,\dots,s$, holds.
\begin{itemize}
\item [\bf(a)]
    Each $\omega_\sigma(\theta)$ is
 even for every variable $\theta_{k+1},\dots,\theta_d$, and, in addition, if $k\ge2$, then
each $\omega_\sigma(\theta)$ is even for some $k-1$ variables from the set $\{\theta_1,\dots,\theta_k\}$.
    \item [\bf(b)]
  Each $\omega_\sigma(\theta)$ is
 even for every variable $\theta_{1},\dots,\theta_k$.
 \item [\bf(c)]
 For every $p=1,\dots,k$, $\ds\sgn \Big(\pa_{\theta_p}\omega_\sigma(\theta)\Big)$
depends only on variable $\theta_p$, and, in addition, if $k\ge3$, then
 each $\omega_\sigma(\theta)$ is
 even for some $k-1$ variables from $\{\theta_1,\dots,\theta_k\}$.
 \end{itemize}
\end{itemize}

For instance, the conditions {\bf(a)}, {\bf(b)}, and {\bf(c)} hold for the nearest neighbor crystal,
see (\ref{omega}).
Under these restrictions on $\omega_\sigma$,
all numbers $ c^l_{p_1\dots p_m}$ in (\ref{4.11}) are equal to zero except for
the case when $m=1$ and $l=p_1\in\{1,\dots,k\}$.
Write
\be\label{4.8'}
c_l\equiv c^l_l=\ds\fr{1}{(2\pi)^{d}}
\sum\limits_{\sigma=1}^s \int_{\T^d}r_\sigma \,\Big|
\frac{\pa\omega_\sigma(\theta)}{\pa\theta_l}\Big|\,d\theta>0,\quad l=1,\dots, k.
\ee
Therefore,
\beqn\label{4.12}
J^l_\infty =\left\{\ba{lll}
-c_l\,\ds
\frac{1}{2^{k}}\sum\limits_{{\bf n}\in{\cal N}^k}
(-1)^{n_{l}} T_{\bf n}=
-c_l\,\frac{1}{2^{k}}\,\,{\sum}'\left(T_{\bf n}\big|_{n_l=2}-T_{\bf n}\big|_{n_l=1}\right),
& l=1,\dots,k,\\
0,& l=k+1,\dots,d,
\ea\right.
\eeqn
where the summation ${\sum}'$ is taken over $n_1,\dots,n_{l-1},n_{l+1},\dots,n_k\in\{1,2\}$.
In particular, if $k=1$, the limiting energy current density is
\be\label{k=1}
{\bf J}_\infty=-\frac12\Big(c_1\left(T_2-T_1\right),0,\dots,0\Big),\quad c_1>0.
\ee
In this case, our model can be considered as a ``system + two reservoirs'',
where by ``reservoirs'' we mean two parts of the crystal consisting
of the particles with $x_1\le -a$ and with $x_1\ge a$, where $a>0$,
and by a ``system''  the remaining (``middle'') part (cf \cite[Sec.3]{SL}).
At $t=0$ the reservoirs are in thermal equilibrium with temperatures $T_1$ and $T_2$.
Therefore, the formula (\ref{k=1})
 corresponds to the Second Law (see, for instance, \cite{BLR},  \cite{SL}),
i.e., the heat flows (on average) from the ``hot reservoir'' to the ``cold'' one.

 If $k=2$, then our model can be considered as a ``system + four reservoirs'',
where reservoirs consist
of the particles with $\{x_1,x_2\le -a\}$,  $\{x_1\le-a,x_2\ge a\}$, $\{x_1\ge a,x_2\le -a\}$,
and $\{x_1,x_2\ge a\}$. The initial states of the reservoirs are
distributed according to Gibbs measures
with corresponding temperatures $T_{11}$,  $T_{12}$,  $T_{21}$, and $T_{22}$.
The formula (\ref{4.12}) becomes
\be\label{4.13}
{\bf J}_\infty=-\frac14
\left(c_1\left(T_{21}-T_{11}+T_{22}-T_{12}\right),
c_2\left(T_{12}-T_{11}+T_{22}-T_{21}\right),0,\dots,0\right),\quad c_1, c_2>0.
\ee
For any $k$, our model is a ``system + $2^k$ reservoirs'',
where ``reservoirs'' are  the crystal particles with position
 $\{x\in\Z^d:(-1)^{n_j}x_j>a\,\,\,\forall j=1,\dots,k\}$,
and  at $t=0$ these reservoirs are assumed to be in thermal equilibrium with temperatures
$T_{\bf n}$, ${\bf n}=(n_1,\dots,n_k)\in{\cal N}^k$.

\begin{remark}
The limiting "kinetic" temperature (average kinetic energy) is
$$
{\bf K}_\infty=\lim_{t\to\infty}\E|\dot u(x,t)|^2=\tr Q^{11}_\infty(x,x)=\tr q^{11}_\infty(0),
$$
by (\ref{1.13}). In the case when $\mu_{\bf n}$ are Gibbs measures with temperatures $T_{\bf n}$,
${\bf K}_\infty$ equals
$$
{\bf K}_\infty=\frac1{(2\pi)^d}\int_{\T^d}\tr \hat q^{11}_\infty(\theta)\,d\theta=
\frac{1}{2^{k}}\sum\limits_{{\bf n}\in{\cal N}^k}T_{\bf n}
\Big(\sum\limits_{\sigma=1}^{s} \frac1{(2\pi)^d} \int_{\T^d}r_\sigma\left(1+
S^{{\rm even}}_{k,{\bf n}}(\omega_\sigma(\theta))\right)d\theta\Big)
$$
by (\ref{4.8}). If all $\omega_\sigma(\theta)$ are even for each $\theta_j$ with $j=1,\dots,k$,
then $\int_{\T^d} S^{{\rm even}}_{k,{\bf n}}(\omega_\sigma(\theta))\,d\theta=0$,
and  ${\bf K}_\infty=n\,2^{-k}\sum\limits_{{\bf n}\in{\cal N}^k}T_{\bf n}$.
For instance, if $k=1$, then ${\bf K}_\infty=n(T_1+T_2)/2$.
\end{remark}

\setcounter{equation}{0}
\section{Convergence of covariance } \label{sec6}

\subsection{Bounds of correlation matrices}\label{sec5}

By $l^p\equiv l^p(\Z^d)\otimes \R^n$, $p,d,n\ge 1$,
 we denote the space of sequences
$f(x)=(f_1(x),\dots,f_n(x))$ endowed with norm
$\Vert f\Vert_{l^p}=\Big(\sum\limits_{x\in\Z^d}|f(x)|^p\Big)^{1/p}$.

\begin{lemma} \label{l4.1}
(i) Let the conditions {\bf S1} and {\bf S2} hold. Then
 for any $\Phi,\Psi\in l^2$,
\be\label{c4.1}
|\langle Q_0(x,y),\Phi(x)\otimes\Psi(y)\rangle|\le
C\Vert\Phi\Vert_{l^2}\Vert\Psi\Vert_{l^2}.
\ee
(ii) Let the conditions {\bf S1}--{\bf S3} hold. Then
$q^{ij}_{\bf n}\in  \ell^1$. Hence,
$\hat q^{ij}_{\bf n}\in  C(\T^d)$, $i,j=0,1$.
\end{lemma}
{\bf Proof}\,
(i) It follows from the bound (\ref{Qh}) that
 $
\sum\limits_{y\in\Z^d}
|Q^{ij}_0(x,y)| \le  \sum\limits_{z\in\Z^d} h(|z|) <\infty.
$
Similarly,
 $\sum\limits_{x\in\Z^d} |Q^{ij}_0(x,y)|\le C<\infty $ for all $y\in\Z^d$.
This implies the bound (\ref{c4.1})  by the Shur lemma.

(ii) The bound  (\ref{Qh})  and condition (\ref{3a}) imply
the same bound for $q^{ij}_{\bf n}(z)$, i.e.,
$|q^{ij}_{\bf n}(z)|\le h(|z|)$, where $r^{d-1}h(r)\in L^1(0,+\infty)$.
Hence,   $q^{ij}_{\bf n}\in l^1$.
\bo

\begin{lemma}\label{lcom}
Let the conditions {\bf E1}--{\bf E3}, {\bf E6}, {\bf S1},  {\bf S2}
hold, and $\alpha<-d/2$. Then  the bound (\ref{20.1}) is true.
\end{lemma}

This lemma can be proved by a same way as in \cite{DKM2}.
We repeat the proof  since some notations and technical bounds obtaining in the proof
we apply in Sec.~\ref{sec5.2} below.\\
{\bf Proof}\,
Note first that
\be\label{E0Q}
\E \Vert  Y(\cdot,t)\Vert^2_\alpha=
\!\sum\limits_{x\in \Z^d}\langle x\rangle ^{2\alpha}
\Big({\rm tr}\,Q_t^{00}(x,x)+{\rm tr}\,Q_t^{11}(x,x)\Big),
\ee
where  $\alpha<-d/2$. Hence,  to prove (\ref{20.1}) it suffices to check that
\be \label{sup}
\sup\limits_{t\in\R} \sup\limits_{x,y\in \Z^d}
\Vert Q_t(x,y)\Vert\le C<\infty.
\ee
Applying the Fourier transform to (\ref{CP}) we obtain
\be\label{CPF}
\dot{\hat Y}(t)=
\hat {\cal A}(\theta)\hat Y(t),\quad t\in\R,
\quad \hat Y(0)=\hat Y_0.
\ee
Here we  denote
$\hat{\cal A}(\theta)=\left(
 \begin{array}{cc}
0 & 1\\
-\hat V(\theta) & 0
\end{array}\right)$, $\theta\in \T^d$.
Therefore, the solution  $\hat Y(\theta,t)$ of (\ref{CPF})
 admits the representation
$\hat Y(\theta,t)=\hat{\cal G}_t(\theta)\hat Y_0(\theta)$
with $\hat{\cal G}_t( \theta):=\exp\Big(\hat{\cal A}(\theta)t\Big)$.
In the coordinate space, we have
\be\label{solGr}
Y(x,t)=\sum\limits_{x'\in\Z^d}{\cal G}_t(x-x')Y_0(x'), \quad x\in\Z^d.
\ee
The  Green function ${\cal G}_t(x)$ has a form (in Fourier transform)
\be\label{hatcalG}
\hat{\cal G}_t( \theta)=
\left( \begin{array}{cc}
 \cos\Omega t &~ \sin \Omega t~\Omega^{-1}  \\
 -\sin\Omega t~\Omega
&  \cos\Omega t\end{array}\right),
\ee
where $\Omega=\Omega(\theta)$ is the Hermitian matrix  defined by (\ref{Omega}).
Then
\be\label{Gtdec}
\hat{\cal G}_t(\theta)=\cos\Omega t\, I+\sin\Omega t\, C(\theta),
\ee
where $C(\theta)$ is defined by (\ref{C(theta)}).
The representation (\ref{solGr}) gives
\beqn\label{5.12}
Q^{ij}_t(x,y)&=&
\E\Big(Y^i(x,t)\otimes Y^j(y,t)\Big)
=\sum\limits_{x',y'\in \Z^d}
\sum\limits_{k,l=0,1}
{\cal G}^{ik}_t(x\!-\!x')Q^{kl}_0(x',y'){\cal G}^{jl}_t(y\!-\!y')\nonumber\\
&=&\langle Q_0(x',y'), \Phi^i_{x}(x',t)\otimes\Phi^j_{y}(y',t)\rangle,
\eeqn
where
$$
\Phi^i_{x}(x',t):=\Big(
{\cal G}^{i0}_t(x-x'),{\cal G}^{i1}_t(x-x')\Big),\quad x'\in\Z^d,\quad i=0,1.
$$
Note that the Parseval identity, (\ref{hatcalG})
and the condition~{\bf E6} imply
$$
\Vert\Phi^i_{x}(\cdot,t)\Vert^2_{l^2}= (2\pi)^{-d}\int_{\T^d}
|\hat\Phi^i_{x}(\theta,t)|^2\,d\theta
=(2\pi)^{-d}\int_{\T^d}\Big(|\hat{\cal G}^{i0}_t(\theta)|^2
+|\hat{\cal G}^{i1}_t(\theta)|^2\Big)\,d\theta\le C_0<\infty.
$$
Then the bound (\ref{c4.1}) gives
\be\label{5.4}
|Q^{ij}_t(x,y)|=|\langle Q_0(x',y'), \Phi^i_{x}(x',t)\otimes
\Phi^j_{y}(y',t)\rangle|
\le C\Vert\Phi^i_{x}(\cdot,t)\Vert_{l^2}\,
\Vert\Phi^j_{y}(\cdot,t)\Vert_{l^2}\le C_1<\infty,
\ee
where the constant $ C_1$  does  not depend on
$x,y\in\Z^d$, $t\in\R$.\bo
\medskip

Proposition \ref{l2.1}  follows  from the bound (\ref{20.1})
by the Prokhorov Theorem \cite[Lemma II.3.1]{VF}
using the method of \cite[Theorem XII.5.2]{VF},
since the embedding ${\cal H}_\alpha\subset {\cal H}_\beta$
is compact if $\alpha>\beta$.

\subsection{Proof of Theorem \ref{tA}}\label{sec5.2}

To prove Theorem \ref{tA}, it suffices to check that for all $\Psi\in {\cal S}$,
\be\label{corfpsi}
{\cal Q}_t(\Psi,\Psi)
\to {\cal Q}_\infty(\Psi,\Psi),\quad t\to \infty.
\ee
 In the cases when $k=0$ and $k=1$,
the convergence (\ref{corfpsi}) was proved in \cite{DKS1} and \cite{DKM1}, respectively.
We derive (\ref{corfpsi}) for any $k\ge1$.
\begin{definition}\label{dC}
(i) The critical set is ${\cal C}:={\cal C}_*\cup{\cal C}_0\cup_{\sigma} {\cal C}_\sigma$
with ${\cal C}_*$ as in Lemma~\ref{lc*} and sets ${\cal C}_0$ and
${\cal C}_\sigma$ defined by (\ref{c0ck}).

(ii)
${\cal S}^0:=\{\Psi\in{\cal S}:
\hat\Psi(\theta)=0\,\,\, \mbox{\rm in a neighborhood of }\,\,{\cal C}\}$.
\end{definition}

Note first that mes${\cal C}=0$.
This fact  can be proved by a similar way as Lemmas~2.2 and 2.3 in \cite{DKS1}
since ${\cal C}\not=\T^d$.
Secondly, we write the scalar product $\langle Y(\cdot,t),\Psi\rangle$ in a form
$$
\langle Y(\cdot,t),\Psi\rangle=\langle Y_0,\Phi(\cdot,t)\rangle,
\quad \mbox{where }\,\,
\Phi(x,t):=F^{-1}_{\theta\to x}[\hat{\cal G}^*_t(\theta)\hat \Psi(\theta)].
$$
Therefore,
\be\label{5.11}
 {\cal Q}_t(\Psi,\Psi)=\E|\langle Y(\cdot,t),\Psi\rangle|^2
 =\langle Q_0(x,y),\Phi(x,t)\otimes\Phi(y,t)\rangle,
\ee
where
the Parseval identity and (\ref{hatcalG}) yield
\beqn\label{esPhi}
\Vert\Phi(\cdot,t)\Vert_{l^2}^2=
(2\pi)^{-d}\int_{\T^d}
\Vert \hat {\cal G}_t^*(\theta)\Vert^2 |\hat\Psi(\theta)|^2 d\theta
\le C \int_{\T^d}\!
\left(1+\Vert V^{-1}(\theta)\Vert\right)|\hat\Psi(\theta)|^2 d\theta
=:C \Vert\Psi\Vert_V^2.\,\,\,
\eeqn
By (\ref{c4.1}), (\ref{5.11}) and (\ref{esPhi}), the uniform bounds hold,
$\sup\limits_{t\in\R} |{\cal Q}_t(\Psi,\Psi)|\le C
\Vert\Psi\Vert_V^2$, $\Psi\in{\cal S}$.
Therefore,
 it suffices to prove the convergence (\ref{corfpsi}) for $\Psi\in{\cal S}^0$ only.
\smallskip

We define a matrix $Q_*(x,y)$, $x,y\in\Z^d$, as follows
\beqn\label{6.8}
Q_*(x,y)\!\!\!&=&\!\!\!\ds\frac1{2^k}\sum\limits_{{\bf n}\in{\cal N}^k}q_{\bf n}(x-y)
\Big(1+(-1)^{n_{1}}\sgn y_{1}\Big)\cdot\dots\cdot\Big(1+(-1)^{n_{k}}\sgn y_{k}\Big)\nonumber\\
\!\!\!\!&=&\!\!\!
\ds\frac1{2^k}\sum\limits_{{\bf n}\in{\cal N}^k}q_{\bf n}(x-y)
\Big[1+\sum\limits_{m=1}^k\sum\limits_{(p_1,\dots,p_m)\in {\cal P}_m(k)}
\!\!\!\!\!(-1)^{n_{p_1}+\dots+n_{p_m}}\sgn y_{p_1}\cdot\dots\cdot\sgn y_{p_m}\Big]\quad\quad
\eeqn
with the matrices $q_{\bf n}(x)$ introduced in the condition~{\bf S3}.
For instance, for $k=1$,
$$
Q_*(x,y)=\frac12\left(q_1(x-y)+q_2(x-y)\right)
+\frac12\left(q_2(x-y)-q_1(x-y)\right)\sgn y_1.
$$
Note that $Q_*(x,y)=q_{\bf n}(x-y)$
in every region $\{(x,y)\in\Z^{2d}:(-1)^{n_1}y_1>0,\,\,\dots,(-1)^{n_k}y_k>0\}$,
 ${\bf n}=(n_1,\dots,n_k)\in{\cal N}^k$.
Denote $Q_r(x,y)=Q_0(x,y)-Q_*(x,y)$. Therefore, the convergence~(\ref{corfpsi})
follows from (\ref{5.11}) and the following proposition.
\begin{pro}\label{p7}
For any $\Psi\in{\cal S}^0$, the following assertions hold.
\beqn\nonumber
\ba{ll}
(a)& \lim\limits_{t\to\infty}\langle Q_*(x,y),\Phi(x,t)\otimes\Phi(y,t)\rangle=
\langle q_\infty(x-y),\Psi(x)\otimes\Psi(y)\rangle.\\~\\
(b)& \lim\limits_{t\to\infty}\langle Q_r(x,y),\Phi(x,t)\otimes\Phi(y,t)\rangle=0.
\ea
\eeqn
\end{pro}

At first, we prove the  auxiliary lemma.
\begin{lemma}\label{l7.1}
Let $q(x)=\left(q^{ij}(x)\right)_{i,j=0,1}$, $x\in\Z^d$, be $2n\times2n$ matrix with  $n\times n$
entries $q^{ij}(x)$ satisfying the bound
$|q^{ij}(x)|\le h(|x|)$, where $r^{d-1}h(r)\in L^1(0,+\infty)$.
Assume that either the condition~{\bf E5} holds or the condition~{\bf E5'}
is fulfilled with the matrices $\hat q^{ij}(\theta)$ instead of $\hat q^{ij}_{\bf n}(\theta)$.
Then for any $\Psi\in{\cal S}^0$,
$$
\lim\limits_{t\to\infty}
\langle q(x-y),\Phi(x,t)\otimes \Phi(y,t)\rangle=\langle {\bf q}_\infty^0(x-y),\Psi(x)\otimes \Psi(y)\rangle,
$$
where
$\hat {\bf q}_\infty^0(\theta)= \sum\limits_{\sigma=1}^s
\Pi_\sigma(\theta) L_1^+\left(\hat q(\theta)\right)\Pi_\sigma(\theta)$, $\theta\in\T^d\setminus {\cal C}_*$.
Moreover, for any $k\in\{1,\dots,d\}$,
$$
\lim\limits_{t\to\infty}
\langle q(x-y)\sgn y_1\cdot\dots\cdot\sgn y_k,\Phi(x,t)\otimes \Phi(y,t)\rangle
=\langle {\bf q}_\infty^k(x-y),\Psi(x)\otimes \Psi(y)\rangle,
$$
where the matrix ${\bf q}_\infty^k(x)$ has a form (in the Fourier transform)
$$
\hat {\bf q}_\infty^k(\theta)= \sum\limits_{\sigma=1}^s
\Pi_\sigma(\theta) {\bf L}_k\left(\hat q(\theta)\right)\Pi_\sigma(\theta)
\sgn(\pa_{\theta_1}\omega_{\sigma}(\theta))\cdot\dots\cdot\sgn(\pa_{\theta_k}\omega_{\sigma}(\theta)),\quad
\theta\in\T^d\setminus {\cal C}_*.
$$
Here
\beqn\label{L}
{\bf L}_k\left(\hat q(\theta)\right)=\left\{
\ba{ll} L_1^+\left(\hat q(\theta)\right),&\mbox{if }\,\, k\,\,\, \mbox{is even},\\
i\,L_2^-\left(\hat q(\theta)\right),&\mbox{if }\,\, k\,\,\, \mbox{is odd},
\ea\right.
\eeqn
where the expressions $L_1^+$ and $L_2^-$ are introduced in (\ref{1.14}).
\end{lemma}
{\bf Proof}\, Using the Fourier transform, we have
\beqn
I_t&:=&\left\langle q(x-y)\sgn y_1\cdot\dots\cdot\sgn y_k,
\Phi(x,t)\otimes \Phi(y,t)\right\rangle\nonumber\\
&=&(2\pi)^{-2d}\int_{\T^{2d}}\Big( F\!\!\!\!_{\scriptsize {\ba{ll}
x\to\theta\\ y\to -\theta'\ea}}\Big[q(x-y)\sgn y_1\cdot\dots\cdot\sgn y_k\Big],
\hat\Phi(\theta,t)\otimes \overline{\hat \Phi(\theta',t)}\,\Big)\,d\theta\, d\theta'.\nonumber
\eeqn
Note that
$F_{y\to \theta}(\sgn y)=
i\,\ds\PV\Big(1/{\tg(\theta/2)}\Big)$, $\theta\in \T^1$, $y\in\Z^1$,
where $\PV$ stands for the Cauchy principal part.
Hence,
\beqn
F\!\!\!\!_{\scriptsize {\ba{ll}
x\to\theta\\ y\to -\theta'\ea}}\Big[q(x-y)\sgn y_1\cdot\dots\cdot\sgn y_k\Big]
=(2\pi)^{d-k}\delta(\tilde \theta-\tilde\theta')\,\hat q(\theta)\times\nonumber\\
\times \,i^k\,
\PV\left(\frac {1}{\tg((\theta_1-\theta_1')/2)}\right)\cdot\dots\cdot
\PV\left(\frac {1}{\tg((\theta_k-\theta_k')/2)}\right),\nonumber
\eeqn
where $\tilde\theta=(\theta_{k+1},\dots,\theta_d)$.
We  choose  a finite partition of unity
\be\label{part}
\sum_{m=1}^M g_m(\theta)=1,\quad \theta\in \supp \hat\Psi,
\ee
where $g_m$ are nonnegative functions from
$C_0^\infty(\T^d)$, which vanish  in a neighborhood of the set
${\cal C}$ introduced in Definition~\ref{dC}~(i).
Using equality $\hat \Phi(\theta,t)=\hat {\cal G}_t^*(\theta)\hat\Psi(\theta)$,
formula (\ref{Gtdec}), decomposition~(\ref{spd'}), and partition~(\ref{part}), we obtain
\beqn\label{7.2}
I_t\!&=&\!(2\pi)^{-d-k}\,i^k\,
\PV\int_{\T^{d+k}}\frac {1}{\tg((\theta_1-\theta_1')/2)}\cdot\dots\cdot
\frac {1}{\tg((\theta_k-\theta_k')/2)}\times\nonumber\\
&&\times\left(\hat {\cal G}_t(\theta)\hat q(\theta)\hat {\cal G}_t^*(\theta'),
\overline{\hat\Psi}(\theta)\otimes \hat \Psi(\theta')\right)\Big|_{\theta'=(\bar\theta',\tilde\theta)}
\,d\bar\theta d\bar\theta' d\tilde\theta\nonumber\\
\!\!&=&\!\!(2\pi)^{-d-k}\,i^k\,\sum\limits_{m,m'}\sum\limits_{\sigma,\sigma'=1}^s
\PV\int_{\T^{d+k}}\!g_m(\theta)g_{m'}(\theta')
\frac{1}{\tg((\theta_1\!-\!\theta_1')/2)}\cdot\dots\cdot
\frac{1}{\tg((\theta_k\!-\!\theta_k')/2)}\!\times\nonumber\\
&&\times\left(\Pi_{\sigma}(\theta)\hat {\cal G}_{t,\sigma}(\theta)\hat q(\theta)
\hat {\cal G}_{t,\sigma'}^*(\theta')\Pi_{\sigma'}(\theta'),
\overline{\hat\Psi}(\theta)\otimes \hat \Psi(\theta')\right)\Big|_{\theta'=(\bar\theta',\tilde\theta)}
\,d\bar\theta d\bar\theta' d\tilde\theta.
\eeqn
Here
\be\label{Gts}
\hat {\cal G}_{t,\sigma}(\theta)
=\cos\omega_\sigma(\theta)t\, I+\sin\omega_\sigma(\theta)t\,C_\sigma(\theta),
\quad C_\sigma(\theta)=\left(\ba{cc}0&1/\omega_\sigma(\theta)\\-\omega_\sigma(\theta)&0\ea\right).
\ee
By Lemma \ref{lc*},
we can choose the supports of $g_m$ so small that the eigenvalues $\omega_\sigma(\theta)$
and the matrices $\Pi_\sigma(\theta)$ are real-analytic functions inside
the $\supp g_m$ for every $m$. (We do not label the functions by the index
$m$ to not overburden the notations.)
Changing variables $\theta'_j\to \xi_j=\theta'_j-\theta_j$, $j=1,\dots,k$,
 in the inner integrals in  the r.h.s. of (\ref{7.2}), we obtain
\beqn\label{6.16}
I_t\!\!&=&\!\!(2\pi)^{-d-k}\,(-i)^k\,\sum\limits_{m,m'}\sum\limits_{\sigma,\sigma'=1}^s
\int_{\T^{d}}\Big(g_m(\theta)\overline{\hat\Psi(\theta)}\Pi_{\sigma}(\theta)
\hat {\cal G}_{t,\sigma}(\theta)\hat q(\theta)\times\nonumber\\
\!\!&&\!\!\times\PV\int_{\T^{k}}
\frac {1}{\tg(\xi_1/2)}\cdot\dots\cdot
\frac {1}{\tg(\xi_k/2)}\,
g_{m'}(\theta')\hat {\cal G}_{t,\sigma'}^*(\theta')\Pi_{\sigma'}(\theta')\hat\Psi(\theta')
\Big|_{\theta'=(\bar\theta+\bar\xi,\tilde\theta)}
d\bar\xi \Big)d\theta.\,\,\,\,\,\,\,\,\,
\eeqn
It follows from Definition \ref{dC}
 that $\partial_{\theta'_j}\omega_{\sigma'}(\theta')\not=0$
for $\theta'\in\supp g_{m'}\subset \supp\hat\Psi$.
The next lemma follows from   \cite[Proposition A.4 i), ii)]{BPT}.
\begin{lemma}\label{auxl}
Let $\chi(\theta)\in C^1(\T^d)$ and
$\partial_{\theta_1}\omega_{\sigma}(\theta)\not=0$ for $\theta\in\supp \chi$. Then
 for $\theta\in\supp \chi$,
 \beqn\label{14.1}
P_{\sigma}(\theta,t)&:=& \PV\int_{\T^1}
\frac{e^{\pm i\omega_{\sigma}(\theta_1+\xi,\tilde\theta)t}}
{\tg(\xi/2)}\chi(\theta_1+\xi,\tilde \theta)\,d\xi\nonumber\\
&=&\pm 2\pi i\,\chi(\theta)\,
e^{\pm i\omega_{\sigma}(\theta)t}\sgn(\pa_{\theta_1} \omega_{\sigma}(\theta))
+o(1)\,\,\,\mbox{as }\,\,t\to+\infty,
\eeqn
where $\tilde \theta=(\theta_2,\dots,\theta_d)$. Moreover,
$\sup\limits_{t\in\R,\,\theta\in \T^d} |P_{\sigma}(\theta,t)|<\infty$.
 Furthermore, using (\ref{Gts}), we have
$$
 \PV\int_{\T^1}
\frac{1}{\tg(\xi/2)}\hat{\cal G}^*_{t,\sigma}(\theta_1+\xi,\tilde\theta)\chi(\theta_1+\xi,\tilde \theta)\,d\xi=
 2\pi \,\chi(\theta)\,C^*_\sigma(\theta)\hat{\cal G}^*_{t,\sigma}(\theta)
\sgn(\pa_{\theta_1} \omega_{\sigma}(\theta))+o(1)
$$
as  $t\to+\infty$.
\end{lemma}

 Applying Lemma \ref{auxl} to the inner  integrals w.r.t.
 $\xi_1,\dots,\xi_k$ in (\ref{6.16}),
we obtain
\beqn\label{6.17}
I_t=(2\pi)^{-d}\,(-i)^k\,\sum\limits_{m}\sum\limits_{\sigma,\sigma'=1}^s
\int_{\T^{d}}g_m(\theta)\Big(\Pi_{\sigma}(\theta)R^k_t(\theta)_{\sigma\sigma'}\Pi_{\sigma'}(\theta),
\hat\Psi(\theta)\otimes\overline{\hat\Psi}(\theta)\Big)\,d\theta+o(1),
\eeqn
where we denote
$R^k_t(\theta)_{\sigma\sigma'}:=
\hat {\cal G}_{t,\sigma}(\theta)\hat q(\theta)\left(C^*_{\sigma'}(\theta)\right)^k
\hat {\cal G}^*_{t,\sigma'}(\theta)$.
Note that $\left(C^*_{\sigma'}(\theta)\right)^k=(-1)^l$ if $k=2l$, and
$\left(C^*_{\sigma'}(\theta)\right)^k=(-1)^lC^*_{\sigma'}(\theta)$ if $k=2l+1$
(with any $l\ge0$).
 Using (\ref{Gts}), we have
\beqn\label{6.18}
R^k_t(\theta)_{\sigma\sigma'}=\left\{\ba{ll}
(-1)^l\sum\limits_{\pm}
\left(\cos\left(\omega^\pm_{\sigma\sigma'}(\theta)t\right)\,
L_1^\mp(\hat q)+\sin\left(\omega^\pm_{\sigma\sigma'}(\theta)t\right)\,
L_2^\pm(\hat q)\right),& k=2l,\\
(-1)^l\sum\limits_{\pm}
\left(\pm\cos\left(\omega^\pm_{\sigma\sigma'}(\theta)t\right)\,
L_2^\pm(\hat q)\mp\sin\left(\omega^\pm_{\sigma\sigma'}(\theta)t\right)\,
L_1^\mp(\hat q)\right),& k=2l+1,
\ea\right.
\eeqn
where $\omega^\pm_{\sigma\sigma'}(\theta)\equiv\omega_\sigma(\theta)\pm\omega_{\sigma'}(\theta)$.
The oscillatory integrals in (\ref{6.17})
with $\omega^\pm_{\sigma\sigma'}(\theta)\not\equiv \const$ vanish as $t\to\infty$
by the Lebesgue--Riemann Theorem, since all integrands in (\ref{6.17}) are summable
by Lemma~\ref{l4.1}~(ii).
Furthermore,  the identities
$\omega^\pm_{\sigma\sigma'}(\theta)\equiv\const_\pm$
 with the $\const_\pm\ne 0$ are impossible by {\bf E5}.
 If we impose the  condition~{\bf E5'} (with $\hat q^{ij}(\theta)$ instead of $\hat q_{\bf n}^{ij}(\theta)$), then
 the case  $\omega^\pm_{\sigma\sigma'}(\theta)\equiv\const_\pm$ (with $\const_\pm\ne 0$)
 is possible. However, in this case,
 $\Pi_{\sigma}(\theta)L_1^\mp(\hat q(\theta))\Pi_{\sigma'}(\theta)\equiv0$
and $\Pi_{\sigma}(\theta)L_2^\pm(\hat q(\theta))\Pi_{\sigma'}(\theta)\equiv0$,
which implies that
$\Pi_{\sigma}(\theta)R^k_t(\theta)_{\sigma\sigma'}\Pi_{\sigma'}(\theta)\equiv0$.
Thus,
 only the integrals with  $\omega^-_{\sigma\sigma'}(\theta)\equiv 0$
contribute to the limit,
since  $\omega^+_{\sigma\sigma'}(\theta)\equiv 0$ would imply
$\omega_{\sigma}(\theta)\equiv\omega_{\sigma'}(\theta)\equiv 0$ which
is impossible by  {\bf E4}. Therefore, using (\ref{6.17}) and (\ref{6.18}), we obtain
$$
I_t=(2\pi)^{-d}\,\sum\limits_{m}\sum\limits_{\sigma=1}^s
\int_{\T^{d}}g_m(\theta)\Big(\Pi_{\sigma}(\theta){\bf L}_k(\hat q(\theta))\Pi_{\sigma}(\theta),
\hat\Psi(\theta)\otimes\overline{\hat\Psi}(\theta)\Big)\,
d\theta+o(1), \quad t\to\infty,
$$
where ${\bf L}_k$ is defined in (\ref{L}).
Lemma \ref{l7.1} is proved. \bo
\medskip

Now Proposition~\ref{p7} (a) follows from
the decomposition (\ref{6.8}), formulas (\ref{1.15})--(\ref{1.14})
and Lemma~\ref{l7.1} with the matrices
$q(x)\equiv q_{\bf n}(x)$.
We prove Proposition \ref{p7} (b) using the methods of \cite[p.140]{BPT} and \cite{DKM1}.
At first, note that
\beqn\label{5.100}
\langle Q_r(x,y),\Phi(x,t)\otimes\Phi(y,t)\rangle
=:\sum\limits_{z\in \Z^d}{\cal F}_t(z),
\eeqn
where
\be\label{calF}
{\cal F}_t(z):=\sum\limits_{y\in \Z^d}
\Big(Q_r(y+z,y),\Phi(y+z,t)\otimes\Phi(y,t)\Big).
\ee
The estimates (\ref{Qh}) and (\ref{3a})
imply the  estimate for $Q_r(x,y)$: $|Q_r(x,y)| \le h(|x-y|)$.
Hence,  the Cauchy--Schwartz inequality and (\ref{esPhi}) give
\beqn
|{\cal F}_t(z)|&\le& \sum\limits_{y\in \Z^d}
\Vert Q_r(y+z,y)\Vert\, |\Phi(y+z,t)|\,|\Phi(y,t)|
\nonumber\\
&\le& h(|z|) \sum\limits_{y\in \Z^d}
|\Phi(y+z,t)|\,|\Phi(y,t)|\le
\!C_1h(|z|) \Vert \Psi\Vert^2_{V},
\eeqn
where $\Vert \Psi\Vert^2_{V}$ is defined in (\ref{esPhi}).
 Since $r^{d-1}h(r)\in L^1(0,+\infty)$, then
\be\label{5.101}
\sum\limits_{z\in \Z^d} |{\cal F}_t(z)|\le
C(\Psi) \sum\limits_{z\in \Z^d}h(|z|)\le C_1<\infty,
\ee
and the series (\ref{5.100})  converges uniformly in $t$.
Therefore, it suffices  to prove that
\be\label{convF}
\lim_{t\to\infty}{\cal F}_t(z)=0 \quad \mbox{for each } z\in \Z^d.
\ee
Let us prove (\ref{convF}).
By (\ref{3.1}) and (\ref{6.8}), $Q_r(x,y)=q_r(\bar x,\bar y,\tilde x-\tilde y)$.
By (\ref{3a}), $\forall \ve>0$
$\exists N\equiv N(\ve)\in\N$ such that $\forall \bar y\in\Z^k$: $|y_j|>N(\ve)$,
$\forall j=1,\dots,k$,
$|q_r(\bar y+\bar z,\bar y,\tilde z)|<\ve$ for any fixed $z=(\bar z,\tilde z)\in\Z^d$.
Hence, by (\ref{esPhi}) and the condition~{\bf E6},
\beqn\label{7.31}
\Big|\sum\limits_{y\in\Z^d:|y_j|>N,\, \forall j=1,\dots,k}\!\!\!
\Big(Q_r(y+z,y),\Phi(y+z,t)\otimes\Phi(y,t)\Big)\Big|
\le \ve \sum\limits_{y\in\Z^d}|\Phi(y,t)|^2 \le \ve\, C(\Psi)\,\,\,\,\,\,\,\,\,\,
\eeqn
uniformly on $t\in\R$. Let us fix $N=N(\ve)$.
Using (\ref{calF}), we obtain
\beqn
|{\cal F}_t(z)|&\le& \ve\, C(\Psi)+C\sum\limits_{j=1}^k\sum\limits_{|y_j|<N}\sum\limits_{y'\in\Z^{d-1}}
\left|\Phi(y+z,t)\right|\left|\Phi(y,t)\right|   
\nonumber\\
&\le& \ve\, C(\Psi)+\sum\limits_{j=1}^k\sum\limits_{|y_j|<N}\sqrt{\sum\limits_{y'\in\Z^{d-1}}
|\Phi(y+z,t)|^2}\sqrt{\sum\limits_{y'\in\Z^{d-1}}|\Phi(y,t)|^2}, 
\nonumber
\eeqn
where, for simplicity, we write $y=(y_j,y')$, $y'=(y_1,\dots,y_{j-1},y_{j+1},\dots,y_d)$.
To prove (\ref{convF}), we fix $j\in\{1,\dots,k\}$ and
verify that for any fixed values of $y_j\in\Z:|y_j|<N$  and  $z\in\Z^d$,
$$
\sum\limits_{y'\in\Z^{d-1}}
|\Phi(y+z,t)|^2\Big|_{y=(y_j,y')}\to0 \quad \mbox{as }\,\,t\to\infty.
$$
Without loss of generality, put $j=1$, $y=(y_1,y')$, $y'=(y_2,\dots,y_d)$.
By the Parseval identity,
\beqn\label{2sum}
\sum\limits_{y'\in \Z^{d-1}}|\Phi(y+z,t)|^2 = (2\pi)^{-2d+2} \int_{\T^{d-1}}
|F_{y'\to \theta'}[\Phi(y+z,t)]|^2\,d\theta'.
\eeqn
It remains to prove that the  integral in the r.h.s. of (\ref{2sum})
tends to zero as $t\to\infty$ for fixed $z\in\Z^d$ and  $|y_1|<N$.
Note first that  the integrand in (\ref{2sum})
satisfies the following uniform bound,
\be\label{7.34}
|F_{y'\to \theta'}[\Phi(y+z,t)]|^2\le
G(\theta'),\,\,\,\,t\ge0,\,\,\,
\mbox{ where }\, G(\theta')\in L^1(\T^{d-1}).
\ee
Indeed, let us rewrite  the function
$F_{y'\to \theta'}[\Phi(y+z,t)]$ in the form
$$
F_{y'\to\theta'}[\Phi(y+z,t)]
=\frac1{2\pi}\int_{\T^1}
e^{-i\theta_1 y_1}e^{-i(\theta, z)}\,\hat\Phi(\theta,t)\,d\theta_1
= \frac1{2\pi}\int_{\T^1}
e^{-i\theta_1 y_1}e^{-i(\theta, z)}\,\hat{\cal G}^*_t(\theta)
\hat\Psi(\theta)\,d\theta_1.
$$
Therefore,
\beqn
|F_{y'\to\theta'}[\Phi(y+z,t)]|^2\!&\le&\!
C\Big(\int_{\T^1}\Vert \hat{\cal G}^*_t(\theta)\Vert\,
|\hat\Psi(\theta)|\,d\theta_1\Big)^2\le
C_1\int_{\T^1}\Vert \hat{\cal G}^*_t(\theta)\Vert^2\,
|\hat\Psi(\theta)|^2\,d\theta_1\nonumber\\
&\le& C_2\int_{\T^1}
\Vert(1+\Vert\hat V^{-1}(\theta)\Vert)
|\hat\Psi(\theta)|^2\,d\theta_1:=G(\theta') \nonumber
\eeqn
and (\ref{7.34}) follows from the condition~{\bf E6}.
Therefore,  it suffices to prove
that the integrand in the r.h.s. of (\ref{2sum})
tends to zero  as $t \to\infty$  for a.a. fixed  $\theta'\in \T^{d-1}$.
We use the  finite partition of unity (\ref{part}),
formulas  (\ref{Gtdec}) and (\ref{spd'})
and split the function $F_{y'\to \theta'}[\Phi(y+z,t)]$
 into the sum of the integrals:
\be\label{7.37}
F_{y'\to\theta'}[\Phi(y+z,t)]=
\sum\limits_m\sum\limits_{\pm}\sum\limits_{\sigma=1}^s\,
\int_{\T^1}
g_m(\theta) e^{-i\theta_1 y_1}e^{-i(\theta,z)}e^{\pm i\omega_\sigma(\theta)t}
a^\pm_\sigma(\theta)\hat \Psi(\theta)\,d\theta_1,
\,\,\,\,\Psi\in{\cal S}^0,
\ee
where
$$
a^\pm_\sigma(\theta):=\frac12\left(\ba{cc}1&\pm i\,\omega_\sigma(\theta)\\
\mp i\,\omega^{-1}_\sigma(\theta)&1\ea\right)\Pi_\sigma(\theta).
$$
The eigenvalues $\omega_\sigma(\theta)$ and the matrices $a^\pm_\sigma(\theta)$
are real-analytic functions inside the $\supp g_m$ for every $m$.
It follows from Definition \ref{dC}~(i) and the conditions~{\bf  E4, E6}
that  mes$\{\theta_1\in \T^1:\,\partial_{\theta_1}\omega_\sigma(\theta)=0\}=0$
for a.a. fixed $\theta'\in \T^{d-1}$.
Hence, the integrals  in (\ref{7.37})
vanish as $t\to\infty$ by  the Lebesgue--Riemann theorem.
 \bo

\setcounter{equation}{0}
\section{Harmonic crystals in the half-space}\label{sec7}
In this section, we consider the dynamics of the harmonic crystals
 in the half-space  $\Z^d_+=\{x\in \Z^d:\,x_1>0\}$, $d\ge 1$,
\be\label{1+}
\ddot u(x,t)=-\sum\limits_{y\in \Z^d_+}\left(V(x-y)-V(x-y_-)\right)
u(y,t),\quad x\in\Z^d_+,\quad t\in\R,
\ee
$y_-:=(-y_1,y_2,\dots,y_d)$,
with zero boundary condition (as $x_1=0$)
\be\label{2+}
u(x,t)|_{x_1=0}=0,
\ee
and with the initial data (as $t=0$)
\be\label{3+}
u(x,0)=u_0(x),\quad \dot u(x,0)=v_0(x),\quad x\in\Z^d_+.
\ee
The matrix $V(x)$ satisfies the conditions {\bf E1}--{\bf E4}.
In addition, we assume that
\be\label{7.4}
V(x_-)=V(x).
\ee
 This condition is fulfilled,
for instance, for the nearest neighbor crystal (\ref{dKG}).
The condition~{\bf E6} imposed on $V(x)$ in Sec.~\ref{sec2.1}
can be weakened as follows.
\begin{itemize}
  \item [{\bf E6'}]
$\ds\int_{\T^d}\sin^2(\theta_1)\Vert \hat V^{-1}(\theta)\Vert\,d\theta<\infty$
\end{itemize}

 Assume that the initial date $Y_0=(u_0,v_0)$ of the problem (\ref{1+})--(\ref{3+})
belongs to the phase space ${\cal H}_{\alpha,+}$, $\alpha\in\R$.
 \begin{definition}
 $ {\cal H}_{\alpha,+}$ is the Hilbert space
of $\R^n\times\R^n$-valued functions  of $x\in\Z^d_+$
 endowed  with  the norm
$ \Vert Y\Vert^2_{\alpha,+}
 = \sum_{x\in\Z^d_+}\langle x\rangle^{2\alpha}\vert Y(x)\vert^2 <\infty$.
\end{definition}

To coordinate the boundary and initial conditions we suppose
that  $u_0(x)=v_0(x)=0$ for $x_1=0$.
\begin{lemma} (see \cite[Corollary 2.4]{D08})
Let the conditions (\ref{7.4}), {\bf E1}, and {\bf E2} hold, and choose some $\alpha\in\R$.
 Then
 for any  $Y_0 \in {\cal H}_{\alpha,+}$, there exists  a unique solution
$Y(t)\in C(\R, {\cal H}_{\alpha,+})$  to the mixed problem (\ref{1+})--(\ref{3+}).
 The operator  $U_+(t):Y_0\mapsto Y(t)$ is continuous
in ${\cal H}_{\alpha,+}$.
\end{lemma}

Below we assume that
  $\alpha<-d/2$ if the condition~{\bf E6} holds and $\alpha<-(d+1)/2$
if the condition~{\bf E6'} holds.
\medskip

We suppose that $Y_0$ is a measurable random function with values in
$\left({\cal H}_{\alpha,+},{\cal B}({\cal H}_{\alpha,+})\right)$
and denote by $\mu^+_0$ a Borel probability measure on ${\cal H}_{\alpha,+}$
giving the distribution of  $Y_0$.
Let $\E_+$ stand for the integral w.r.t.  $\mu^+_0$,
and denote by $Q^+_0(x,y)$ the initial covariance of $\mu^+_0$,
$$
Q^+_0(x,y)=\E_+\left(Y_0(x)\otimes Y_0(y)\right)
\equiv \int\left(Y_0(x)\otimes Y_0(y)\right)\,\mu_0^+(dY_0),\quad x,y\in\Z^d_+.
$$
In particular, $Q^+_0(x,y)=0$ for $x_1=0$ or $y_1=0$.
We assume that $\mu^+_0$ satisfies the conditions~{\bf S1} and {\bf S2}
stated in Sec.~\ref{sec2.1}.
The condition~{\bf S3} needs in some modification.
\begin{itemize}
\item [{\bf S3}]
Choose some $k\in\{1,\dots,d\}$.
The initial covariance $Q^+_0(x,y)$  has a form
\be\label{3.1+}
Q^+_0(x,y)=q^+_0(\bar x,\bar y,\tilde x-\tilde y),\quad x,y\in\Z^d_+,
\ee
where $x=(\bar x,\tilde x)$,
$\bar x=(x_1,\dots,x_k)$, $\tilde x=(x_{k+1},\dots,x_d)$.
Write (cf (\ref{N}))
$$
{\cal N}^k_+:=\{{\bf n}=(n_1,n_2,\dots,n_k),\,\,\, n_1=2,\,\,\,n_j\in\{1,2\}\,\,\,\mbox{for all }\,\, j=2,\dots,k\}.
$$
Suppose that  $\forall\ve>0$  $\exists N(\ve)\in\N$
such that for any
$\bar y=(y_1,\dots,y_k)\in\Z^k$: $y_1>N(\ve)$ and $(-1)^{n_j}y_j>N(\ve),\,\forall j=2,\dots,k$,
the following bound holds (cf (\ref{3a}))
\be\label{3a+}
\left|q^+_0(\bar y+\bar z,\bar y,\tilde z)-q_{\bf n}(z)\right|<\ve
\quad\mbox{for any fixed }\,z=(\bar z,\tilde z)\in\Z^d.
\ee
Here
$q_{\bf n}(z)$, ${\bf n}\in{\cal N}^k_+$, are  the correlation matrices of some
translation-invariant measures $\mu_{\bf n}$
with zero mean value  in ${\cal H}_\alpha$.
\end{itemize}

In particular, if $k=1$, then $Q^+_0(x,y)=q^+_0(x_1,y_1,\tilde x-\tilde y)$,
$\tilde x=(x_{2},\dots,x_d)$, and (cf (\ref{1.7'}))
\be\label{7.6}
q^+_0(y_1+z_1,y_1,\tilde z)\to q_2(z)\quad \mbox{as }\,\,y_1\to+\infty.
\ee

\begin{example}
{\rm The example of $\mu_0^+$ satisfying the conditions {\bf S1}--{\bf S3}
can be constructed by a similar way as for $\mu_0$ in Sec.~\ref{sec3}.
Indeed, define a  Borel probability measure
 $\mu_0$ as a distribution of the random function (cf (\ref{rf-k}))
$$
Y_0(x)= \sum_{{\bf n}\in{\cal N}^k_+}\ov\zeta_{{\bf n}}(\bar x)Y_{\bf n}(x),
\quad x=(\bar x,\tilde x)\in \Z^d_+, \quad \bar x=(x_1,\dots, x_k),
\quad \tilde x=(x_{k+1},\dots,x_d),
$$
where $\ov\zeta_{\bf n}(\bar x)=\zeta_{2}( x_1)\zeta_{n_2}( x_2)\cdot\dots\cdot\zeta_{n_k}( x_k)$
for $ \bar x=(x_1,\dots,x_k)$,
 the functions $\zeta_n$ are defined in (\ref{zeta}),
$ Y_{\bf n}(x)$, ${\bf n}\in{\cal N}^k_+$, are Gaussian independent vectors in ${\cal H}_{\alpha,+}$
with distributions $\mu_{\bf n}$.
}\end{example}

 We define $\mu^+_t$, $t\in\R$, as a Borel probability measure in ${\cal H}_{\alpha,+}$
 which gives the distribution of the random solution $Y(t)$,
$\mu^+_t(B)=\mu_0(U_+(-t)B)$, $ B\in {\cal B}({\cal H}_{\alpha,+})$, $t\in \R$.
Denote by $Q_t^+(x,y)=\int \left(Y(x)\otimes Y(y)\right)\,\mu_t^+(dY_0)$, $x,y\in\Z^d_+$,
the covariance of $\mu_t^+$.
The mixing condition~{\bf S4} (see Sec.~\ref{sec2.3})
for $\mu^+_0$ is formulated  as for the measure
$\mu_0$ but with sets ${\cal A}$ and ${\cal B}$ from $\Z^d_+$ instead of $\Z^d$.
\medskip

Introduce the limiting correlation matrix $Q^+_\infty(x,y)$. It has a form
\be\label{1.15+}
Q^+_\infty(x,y)=q^+_\infty(x-y)-q^+_\infty(x-y_-)-q^+_\infty(x_--y)+
q^+_\infty(x_--y_-),\quad x,y\in\Z^d_+.
\ee
Here $q^+_\infty(x)$ is defined as $q_\infty(x)$ (see formulas
 (\ref{1.15})--(\ref{C(theta)})) but with ${\cal N}_+^k$ instead of ${\cal N}^k$.
 For example,  if $k=1$, then $\hat q^+_\infty(\theta)$ has a form (\ref{1.15})
 with matrices (cf (\ref{2.22}))
 $$
 {\bf M}^+_{1,\sigma}(\theta)=\frac12 L_1^+\left(\hat q_2(\theta)\right),\quad
 {\bf M}^-_{1,\sigma}(\theta)=\frac12 L_2^-\left(\hat q_2(\theta)\right)
 \sgn  \left(\partial_{\theta_1}\omega_\sigma(\theta)\right),
 $$
 where $\hat q_2(\theta)$ is the Fourier transform of the matrix $q_2(z)$ introduced  in (\ref{7.6}).
\begin{theorem}\label{tA+}
Assume that
  $\alpha<-d/2$ if the condition~{\bf E6} holds and $\alpha<-(d+1)/2$
if the condition~{\bf E6'} holds. Then the following assertions are valid.\\
(i)  Let the conditions (\ref{7.4}), {\bf E1}--{\bf E4}, {\bf E5'}, {\bf E6'},
and   {\bf S1}--{\bf S3} be fulfilled.
 Then for any $x,y\in\Z^d$,
$Q_t^+(x,y)\to Q_\infty^+(x,y)$ as $t\to\infty$,
where $Q_\infty^+$ is defined in (\ref{1.15+}).\\
(ii)  Let the  conditions (\ref{7.4}), {\bf E1}--{\bf E3}, {\bf E4'}, {\bf E5'}, {\bf E6'},
{\bf S1}, {\bf S3},  and {\bf S4} be fulfilled.
 Then the measures  $\mu_t$ weakly converge in  the Hilbert space
 ${\cal H}_{\alpha,+}$  as $t\to \infty$.
The limiting measure $ \mu^+_\infty $ is a Gaussian
measure on ${\cal H}_{\alpha,+}$
with the covariance $Q_\infty^+(x,y)$ defined in (\ref{1.15+}).
\end{theorem}

The second assertion of Theorem \ref{tA+} can be
proved by a similar way as Theorem \ref{tB}.
The proof of the first assertion has some features in comparison with
the proof of Theorem \ref{tA}, see Sec.~\ref{sec7.1} below.

\subsection{The proof}\label{sec7.1}
\begin{lemma} (cf Lemma \ref{lcom})
Let the conditions (\ref{7.4}), {\bf E1}--{\bf E3}, {\bf E6'}, {\bf S1}, and {\bf S2} be fulfilled.
Then the uniform bound holds, $\sup_{t\in\R}\E_+\left(\Vert Y(t)\Vert^2_{\alpha,+}\right)<\infty$.
\end{lemma}
{\bf Proof}\,
By $l^2_+\equiv l^2(\Z^d_+)\otimes \R^n$, $d,n\ge 1$,
 we denote the Hilbert space of sequences
$f(x)=(f_1(x),\dots,f_n(x))$ endowed with norm
$\Vert f\Vert_{l^2_+}=\sqrt{\sum\limits_{x\in\Z^d_+}|f(x)|^2}$.
Let $\langle\cdot,\cdot\rangle_+$ stand for the inner product in $\ell^2_+$
(or in $\ell^2_+\times\ell^2_+$).
At first, by the conditions~{\bf S1} and {\bf S2}, we have (cf (\ref{c4.1}))
\be\label{7.7}
|\langle Q^+_0(x,y),\Phi(x)\otimes\Psi(y)\rangle_+|\le
C\Vert\Phi\Vert_{l^2_+}\Vert\Psi\Vert_{l^2_+}\quad\mbox{ for any }\,\,
 \Phi,\Psi\in \ell^2_+\times\ell^2_+.
\ee
Second, the solutions of the problem (\ref{1+})--(\ref{3+}) has a form
\be\label{7.5}
Y(x,t)=\sum\limits_{x'\in\Z^d_+}{\cal G}_{t,+}(x,x')Y_0(x'),\quad\mbox{where }\,\,
{\cal G}_{t,+}(x,x')={\cal G}_{t}(x-x')-{\cal G}_{t}(x-x'_-),
\ee
with ${\cal G}_{t}(x)$ defined in (\ref{hatcalG}).
Similarly to (\ref{5.12}), we have
\beqn\label{7.8}
\left(Q^{+}_t(x,y)\right)^{ij}=
\left\langle Q^+_0(x',y'), \Phi^i_{x}(x',t)\otimes\Phi^j_{y}(y',t)\right\rangle_+,
\quad x,y\in\Z^d_+,
\eeqn
where
$\Phi^i_{x}(x',t):=\left(
{\cal G}^{i0}_{t,+}(x,x'),{\cal G}^{i1}_{t,+}(x,x')\right)$,  $i=0,1$.
By the Parseval identity, formula~(\ref{hatcalG}),  the condition~{\bf E6'}
and Fei\'er's theorem, we have
\beqn\label{7.9}
\Vert\Phi^i_{x}(\cdot,t)\Vert^2_{l^2}&=& (2\pi)^{-d}\int_{\T^d}
\left|\hat\Phi^i_{x}(\theta,t)\right|^2\,d\theta
=(2\pi)^{-d}4\int_{\T^d}\sin^2(\theta_1 x_1)\left(|\hat{\cal G}^{i0}_t(\theta)|^2
+|\hat{\cal G}^{i1}_t(\theta)|^2\right)\,d\theta\nonumber\\
&\le& \int_{\T^d}\sin^2(\theta_1 x_1)\left(C_1+C_2\Vert\hat V^{-1}(\theta)\Vert\right)\,d\theta
\le C_3+C_4|x_1|,
\eeqn
where constants $C_3$ and $C_4$ do not depend on $t\in\R$ and $x\in\Z^d$, and
 $C_4=0$ if the condition~{\bf E6} holds.
 Hence, (\ref{7.7}), (\ref{7.8}) and (\ref{7.9}) imply
$$
|\left(Q^{+}_t(x,y)\right)^{ij}|\le C \Vert \Phi^i_{x}(\cdot,t)\Vert_{\ell^2_+}
\Vert\Phi^j_{y}(\cdot,t)\Vert_{\ell^2_+}\le C\sqrt{C_3+C_4|x_1|}\sqrt{C_3+C_4|y_1|},
\quad x,y\in\Z^d_+.
$$
Therefore, the choice of $\alpha$ implies the following bound
\beqn\nonumber
\E_+ \Big(\Vert  Y(\cdot,t)\Vert^2_{\alpha,+}\Big)&=&
\sum\limits_{x\in \Z^d_+}\langle x\rangle^{2\alpha}\,
{\rm tr}\,\left(\left(Q^+_t(x,x)\right)^{00}+
\left(Q^+_t(x,x)\right)^{11}\right)\nonumber\\
&\le& C
\sum\limits_{x\in \Z^d_+}\langle x\rangle^{2\alpha}\left(C_3+C_4|x_1|\right)<\infty.
\nonumber
\bo
\eeqn
{\bf Proof of Theorem \ref{tA+} (i)}:
At first, using (\ref{7.5}), we decompose the covariance $Q^+_t(x,y)$ into a sum
of four terms:
$$
Q^+_t(x,y)=\sum\limits_{x',y'\in\Z^d_+}{\cal G}_{t,+}(x,x')Q^+_0(x',y')
{\cal G}^T_{t,+}(y,y')
=R_t(x,y)-R_t(x,y_-)-R_t(x_-,y)+R_t(x_-,y_-),
$$
where $(\,)^T$ denotes matrix transposition,
$$
R_t(x,y):=\sum\limits_{x',y'\in\Z^d_+}
{\cal G}_t(x-x')Q^+_0(x',y'){\cal G}^T_t(y-y').
$$
Therefore, Theorem \ref{tA+} (i) follows from the following convergence
\be\label{5.6}
R_t(x,y)\to q^+_\infty(x-y)\quad \mbox{as }\,t\to\infty,\quad
x,y\in\Z^d.
\ee
To prove (\ref{5.6}), let us define
$ \bar Q^+_0(x,y)$ to be equal to $Q_0^+(x,y)$ for $x,y\in\Z^d_+$,
and to $0$ otherwise.
Denote by  $Q^+_*(x,y)$ the matrix which is defined as $Q_*(x,y)$ (see (\ref{6.8}))
but with the summation over ${\cal N}^k_+$ instead of ${\cal N}^k$.
Put $Q^+_r(x,y)=\bar Q_0^+(x,y)-Q_*^+(x,y)$.
Then (\ref{5.6}) follows from the following assertions. For any $x,y\in\Z^d$,
\beqn\nonumber
\ba{lll}
\sum\limits_{x',y'\in\Z^d}
{\cal G}_t(x-x')Q^+_*(x',y'){\cal G}^T_t(y-y')&\to&
 q_\infty^+(x-y),\quad t\to\infty,\\
\sum\limits_{x',y'\in\Z^d}
{\cal G}_t(x-x')Q^+_r(x',y'){\cal G}^T_t(y-y')&\to& 0,\quad t\to\infty.
\ea
\eeqn
The proof of these assertions are similarly to the proof of Proposition \ref{p7}.\bo

\subsection{Energy current in the half-space}

Here we calculate the limiting energy current density ${\bf J}_{+,\infty}=(J^1_{+,\infty},\dots,J^d_{+,\infty})$.
\begin{lemma}\label{l7.2}
If $d=1$, then ${\bf J}_{+,\infty}=0$.
If $d\ge 2$,
the coordinates of the energy current density ${\bf J}_{+,\infty}\equiv {\bf J}_{+,\infty}(x_1)$,
$x_1\ge0$, are
\beqn\label{7.10}
J^1_{+,\infty}(x_1)\equiv0,\,\,\,
 J^l_{+,\infty}(x_1) = -\fr{2i}{(2\pi)^{d}}
\int_{\T^d} \!\sin^2(\theta_1x_1)\tr\left[\left(\hat q^+_\infty (\theta)\right)^{10}
 \partial_{\theta_l} \hat V(\theta)\right]\!d\theta, \,\,\, l=2,\dots,d,\,
\eeqn
with $q_\infty^+$ from (\ref{1.15+}). In particular, ${\bf J}_{+,\infty}(0)=0$.
\end{lemma}

To prove (\ref{7.10}), we first
derive formally the expression of the energy current for
the finite energy solutions $u(x,t)$.
For the region $\Omega_l:=\{x\in\Z^d_+:\,x_l\ge 0\}$, $l\ge1$,
we define the energy in  $\Omega_l$  as
$$
{\cal E}_{+}^{l}(t):=\frac12\sum\limits_{x\in\Omega_l}
\Big\{|\dot u(x,t)|^2+\sum\limits_{y\in\Z^d_+}
\Big(u(x,t),(V(x-y)-V(x-y_-))u(y,t)\Big)\Big\}.
$$
Then, using Eqn (\ref{1+}), (\ref{7.4}) and {\bf E2},  we obtain
$$
\dot {\cal E}_{+}^{1}(t)=0,\quad
\dot {\cal E}_{+}^{l}(t)=\sum\limits_{x'\in\Z^d_+}J^l_+(x',t),\quad l=2,\dots,d.
$$
Here $J^l_+(x',t)$ stands for the  energy current density
in the direction $e_l=(0,\delta_{l2},\dots,\delta_{ld})$,
$$
 J^l_+(x',t):=\fr12\sum\limits_{y'\in\Z^d_+}
\Big\{\sum\limits_{m\le-1,\,p\ge 0} A^l_{mp}(x',y',t)-\sum\limits_{m\ge0,\,p\le -1} A^l_{mp}(x',y',t)\Big\},
$$
where $A^l_{mp}(x',y',t):=\Big(\dot u(x,t),\left(V(x-y)-V(x-y_-)\right)u(y,t)\Big)$
for  $x\equiv x'+me_l$, $y\equiv y'+pe_l$,
 $x',y'\in\Z^d_+$ with $x'_l=y'_l=0$, $l=2,\dots,d$.

Let $u(x,t)$ be the random solution to problem (\ref{1+})--(\ref{3+}) with
the initial measure $\mu^+_0$ satisfying {\bf S1}--{\bf S3}.
Using Theorem \ref{tA+} (i), we have
$$
\E_+\left(J^l_+(x',t)\right)\to   J^l_{+,\infty}
:=\fr12 \sum\limits_{y'\in \Z^d_+}
\Big\{\sum\limits_{m\le-1,\,p\ge 0}B^l_{mp}(x',y')-\sum\limits_{m\ge0,\,p\le -1}B^l_{mp}(x',y')\Big\}
\quad \mbox{as }\,\,t\to\infty,
$$
where $B^l_{mp}(x',y'):=\tr\Big[Q^{10}_\infty (x,y)
\left(V^T(x-y)-V^T(x-y_-)\right)\Big]$,
 $x\equiv x'+me_l$, $y\equiv y'+pe_l$,
 $x',y'\in\Z^d_+$ with $x'_l=y'_l=0$.
Applying (\ref{1.15+}), we obtain
$$
J^l_{+,\infty}=-\frac12\tr\sum\limits_{y\in \Z^d} y_l\left(\left(q_\infty^+(x'+y)\right)^{10}
-\left(q_\infty^+(x'_-+y)\right)^{10}\right)
\left(V^T(x'+y)-V^T(x'+y_-)\right).
$$
Using the equality $V(x)=V(x_-)$ and applying the Fourier transform,
 we obtain (\ref{7.10}).
Lemma~\ref{l7.2} is proved. \bo
\medskip

Let $\mu_{\bf n}=g_{\beta_{\bf n}}$, ${\bf n}\in{\cal N}^k_+$, be the Gibbs measures
constructed in Sec.~\ref{sec4.2} with temperatures $T_{\bf n}>0$.
The correlation matrices of $\mu_{\bf n}$ are
 $q_{\bf n}(x-y)\equiv q_{\beta_{\bf n}}(x-y)$, see (\ref{4}).
We impose, in addition, the condition~(\ref{E6'}) on the matrix $V$,
which implies the bound (\ref{ft}) for $q^{00}_{\bf n}$.
Then, the condition {\bf S2} is fulfilled. Since
$$
\left(\hat q^+_{\infty}(\theta)\right)^{10}
=-i\,\sum\limits_{\sigma=1}^s \omega_\sigma^{-1}(\theta)
\Pi_\sigma(\theta)\Big(\frac1{2^k}\sum\limits_{{\bf n}\in{\cal N}^k_+}T_{\bf n}
S^{{\rm odd}}_{k,{\bf n}}(\omega_\sigma(\theta))\Big),
$$
where the function $S^{{\rm odd}}_{k,{\bf n}}(\omega_\sigma)$
 is defined  in (\ref{Sodd}),
then for $l=2,\dots,d$ (cf (\ref{mecd}))
\beqn\label{mecd+}
 J^l_{+,\infty}(x_1) &=&-\fr{4}{(2\pi)^{d}}
\sum\limits_{\sigma=1}^{s}\int_{\T^d} r_\sigma \sin^2(\theta_1x_1)
\Big(\frac1{2^k}\sum\limits_{{\bf n}\in{\cal N}^k_+}T_{\bf n}
S^{{\rm odd}}_{k,{\bf n}}(\omega_\sigma(\theta))\Big)\,
\frac{\pa\omega_{\sigma}(\theta)}{\pa\theta_l}\,d\theta
\nonumber\\
&=&-\sum\limits_{{\rm odd}\, m\in\{1,\dots, k\}}
\sum\limits_{(p_1,\dots,p_m)\in {\cal P}_m(k)}c^l_{p_1\dots p_m}(x_1)\,
 \frac{1}{2^{k-1}}\sum\limits_{{\bf n}\in{\cal N}^k_+}
(-1)^{n_{p_1}+\dots+n_{p_m}} T_{\bf n},
\eeqn
where the functions $c^l_{p_1\dots p_m}(x_1)$, $x_1\in\Z^1_+$, are defined as follows
(cf (\ref{4.11}))
\be\label{4.11+}
c^l_{p_1\dots p_m}(x_1):=
\fr{2}{(2\pi)^{d}}\sum\limits_{\sigma=1}^{s}\int_{\T^d}r_\sigma \sin^2(\theta_1x_1)
\sgn \left( \frac{\partial\omega_\sigma(\theta)}{\partial\theta_{p_1}}\right)\cdot\dots
 \cdot
\sgn \left(\frac{\partial \omega_\sigma(\theta)}{\partial \theta_{p_m}}\right)
\frac{\pa\omega_{\sigma}(\theta)}{\pa\theta_l}\,d\theta.
\ee
Write
\be\label{cl}
c_l(x_1)\equiv c^l_l(x_1)=\ds\fr{2}{(2\pi)^{d}}
\sum\limits_{\sigma=1}^s \int_{\T^d}r_\sigma \sin^2(\theta_1x_1)\Big|
\frac{\pa\omega_\sigma(\theta)}{\pa\theta_l}\Big|\,d\theta>0,\quad l=2,\dots, k.
\ee
Applying the condition {\bf SC} to $\omega_\sigma(\theta)$, we obtain
\beqn\label{7.17}
J^l_{+,\infty}(x_1) =\left\{\ba{ll}
\ds
-c_l(x_1)\frac{1}{2^{k-1}}\,\,{\sum\limits_{{\bf n}\in{\cal N}_+^k}}'
\Big(T_{\bf n}\big|_{n_l=2}-T_{\bf n}\big|_{n_l=1}\Big),
& l=2,\dots,k,\\
0,& l=1,\,l=k+1,\dots,d.
\ea\right.
\eeqn
where the summation ${\sum}'$ is taken over $n_2,\dots,n_{l-1},n_{l+1},\dots,n_k\in\{1,2\}$.
Using the formula $2\sin^2(\theta_1 x_1)=1-\cos(2\theta_1x_1)$
and the Lebesgue--Riemann theorem, we see that
 $c_l(x_1)\to c_l$ as $x_1\to+\infty$, where
the positive constant $c_l$ is defined in (\ref{4.8'}). Hence, for $l=2,\dots,k$,
\be\label{7.18}
J^l_{+,\infty}(x_1)\to -c_l\,
\frac{1}{2^{k-1}}\,\,{\sum}'\left(T_{\bf n}\big|_{n_l=2}-T_{\bf n}\big|_{n_l=1}\right)
\quad \mbox{as }\,x_1\to+\infty.
\ee

Consider the particular cases of (\ref{mecd+}) and (\ref{7.17}).
\begin{example}\label{example7.7}
Let $k=1$ and $\mu^+_0$ satisfy the condition~{\bf S3}
with a Gibbs measure $\mu_2\equiv g_{\beta}$, $\beta=1/T_2$.
For instance, the initial datum $Y_0$ has a form
$Y_0(x)=\zeta_2(x_1)Y_2(x)$, where $\zeta_2$ is defined in (\ref{zeta}),
$Y_2$ has the Gibbs distribution $g_{\beta}$.
Hence, $J^1_{+,\infty}\equiv0$, and
$$
J^l_{+,\infty}(x_1)=-
\fr{2T}{(2\pi)^{d}}\sum\limits_{\sigma=1}^{s}\int_{\T^d}r_\sigma \sin^2(\theta_1x_1)
\sgn \left(\frac{\partial \omega_\sigma(\theta)}{\partial \theta_{1}}\right)
\frac{\pa\omega_{\sigma}(\theta)}{\pa\theta_l}\,d\theta,
\quad l=2,\dots,d.
$$
Suppose that the eigenvalues $\omega_\sigma(\theta)$ satisfy the
symmetry conditions {\bf SC}.
Then ${\bf J}_{+,\infty}(x_1)=0$ for any $x_1\ge0$.
\end{example}
\begin{example}
Let $k=2$ and $\mu^+_0$ satisfy the condition~{\bf S3}
with Gibbs measures $\mu_{\bf n}\equiv g_{\beta_{\bf n}}$, $\beta_{\bf n}=1/T_{\bf n}$,
${\bf n}=(n_1,n_2)\in{\cal N}^2_+=\{(2,1);(2,2)\}$.
For instance, the initial datum $Y_0$ is of a form
$$
Y_0(x)=\zeta_2(x_1)\Big(\zeta_{1}(x_2)Y_{21}(x)+\zeta_{2}(x_2)Y_{22}(x)\Big),
\quad x\in\Z^d_+,
$$
where $\zeta_{n}(x)$ is defined in (\ref{zeta}), $Y_{21}(x)$ and $Y_{22}(x)$
are independent vectors in ${\cal H}_{\alpha}$ with Gibbs distributions
$\mu_{21}$ and $\mu_{22}$, corresponding positive temperatures
$T_{21}$ and $T_{22}$, respectively.
Therefore, $J^1_{+,\infty}(x_1)\equiv0$, and
\beqn
J^l_{+,\infty}(x_1)&=&-
\fr{1}{(2\pi)^{d}}\sum\limits_{\sigma=1}^{s}
\int_{\T^d}r_\sigma\sin^2(\theta_1x_1)
\Big[{\rm sign} \left(\frac{\partial \omega_\sigma(\theta)}{\partial \theta_{1}}\right)
(T_{21}+T_{22})\nonumber\\
&&+ \sgn \left(\frac{\partial \omega_\sigma(\theta)}{\partial \theta_{2}}\right)
(T_{22}-T_{21})\Big]
\frac{\pa\omega_{\sigma}(\theta)}{\pa\theta_l}\,d\theta, \quad l=2,\dots,d.
\nonumber
\eeqn
Under the additional conditions~{\bf SC}  on the eigenvalues $\omega_\sigma(\theta)$,
we obtain
$$
{\bf J}_{+,\infty}(x_1)=-\frac12\Big(0,c_2(x_1)(T_{22}-T_{21}),0,\dots,0\Big)
$$
with $c_2(x_1)$ introduced in (\ref{cl}).
Note that (cf (\ref{k=1}), (\ref{4.13}))
$$
{\bf J}_{+,\infty}(x_1)\to -\frac12\Big(0,c_2(T_{22}-T_{21}),0,\dots,0\Big)
\quad \mbox{as }\,x_1\to+\infty,
$$
where the positive constant $c_2$ is defined in (\ref{4.8'}).
\end{example}
\begin{remark}
{\rm In \cite{D17},
 we consider the 1D chain  of harmonic oscillators on the half-line with
 {\em nonzero} boundary condition and
 study the following initial boundary value problem:
$$
\left\{\ba{lr}
\ddot u(x,t)=(\Delta_L-m^2) u(x,t),\quad\quad x\ge1,\quad t>0,\\
\ddot u(0,t)=-\kappa u(0,t)-m^2u(0,t)-\gamma\dot u(0,t)+u(1,t)-u(0,t),
\quad t>0,\\
u(x,0)=u_0(x),\quad \dot u(x,0)=v_0(x),\quad x\ge0.
\ea\right.$$
Here $u(x,t)\in\R$, $m\ge0$,  $\gamma\ge0$,
 $\Delta_L$ denotes the second derivative on $\Z$.
We impose some restrictions on the coefficients $m,\kappa,\gamma$
of the system. In particular,
if $\gamma>0$, then either $m>0$ or  $\kappa>0$.
If $\gamma=0$, then $\kappa\in(0,2)$.
We obtain the results similar to  (\ref{corf}) and (\ref{1.8i}).
Furthermore, the limiting energy current at the origin equals
$J_\infty:=-\gamma \lim\limits_{t\to\infty}\E\left(\dot u(0,t)\right)^2$.
Hence, in the case when $\gamma>0$, $J_\infty\not=0$ (cf Example \ref{example7.7})
if $\int(Y^1(0))^2\,\mu_\infty(dY)\not=0$
(the limit measures $\mu_\infty$ satisfying the last condition are constructed in \cite{D17}).
}\end{remark}

{\bf ACKNOWLEDGMENT}
\medskip

 This work was supported partly by the research grant of RFBR
(Grant no. 18-01-00524).



\begin{thebibliography}{99}

\bibitem{BPT}
Boldrighini, C., Pellegrinotti, A., and Triolo, L.,
``Convergence to stationary states for infinite harmonic systems,''
J. Stat. Phys. {\bf 30}, 123--155 (1983).

\bibitem{BLR}
Bonetto, F., Lebowitz, J.L., and Rey-Bellet, L.,
Fourier law: a challenge to theorists, p.128-150
in:  Fokas, A. (ed.) et al., Mathematical physics 2000.
International congress, London, GB,
Imperial College Press, London, 2000. ArXiv: math-ph/0002052.

\bibitem{BLL}
Bonetto, F., Lebowitz, J.L., and Lukkarinen, J.,
``Fourier's law for a harmonic crystal with self-consistent stochastic reservoirs,''
 J. Stat. Phys. {\bf 116} (1/4), 783--813 (2004).

\bibitem{CaLeb}
Casher, A., and Lebowitz, J.L.,
``Heat flow in regular and disordered harmonic chains,''
 J. Math. Phys. {\bf 12} (8), 1701--1711 (1971).


\bibitem{CFS} Cornfeld, I.P., Fomin, S.V., and Sinai, Ya.G.,
  {\it Ergodic Theory} (Springer, New York, 1981).

\bibitem{DS} Dobrushin, R.L., and Suhov, Yu.M.,
``On the problem of the mathematical
foundation of the Gibbs postulate in classical statistical mechanics,''
 in {\it Mathematical Problems in Theoretical Physics},
Lecture Notes in Physics, vol. 80
(Springer-Verlag, Berlin, 1978), p. 325--340.


\bibitem {DKS1} Dudnikova, T.V., Komech, A.I., and Spohn, H.,
 ``On the convergence to statistical equilibrium for harmonic crystals,''
J. Math. Phys. {\bf 44}, 2596--2620 (2003).

\bibitem{DKM1}  Dudnikova, T.V., Komech, A.I., and Mauser, N.J.,
``On two-temperature problem for harmonic crystals,''
  J. Stat. Phys. {\bf 114}, no.3/4, 1035--1083 (2004).

\bibitem{DKM2}
 Dudnikova, T.V., and Komech, A.I.,
``On a two-temperature problem for the Klein--Gordon equation,''
 Theory Prob. Appl. {\bf 50} (4), 582--611 (2006).

\bibitem{D08}
Dudnikova, T.V.,
``On the asymptotical normality of statistical solutions for harmonic crystals in half-space,''
 Russian J. Math. Phys.  {\bf 15} (4), 460--472 (2008).

\bibitem{D17} Dudnikova, T.V.,
``On convergence to equilibrium for one-dimensional chain of harmonic oscillators on the half-line,''
 J. Math. Phys. {\bf 58} (4), 043301 (2017).

\bibitem{EPR1}
Eckmann, J.-P., Pillet C.-A., and Rey-Bellet, L.,
``Non-equilibrium statistical mechanics of
anharmonic chains coupled to two heat baths at different temperatures,''
Commun. Math. Phys. {\bf 201}, 657--697 (1999).

\bibitem{EPR2}  Eckmann, J.-P., Pillet C.-A., and Rey-Bellet, L.,
``Entropy production in nonlinear, thermally driven Hamiltonian systems,''
J. Stat. Phys. {\bf 95} (1/2), 305--331 (1999).

\bibitem{FL}
Fidaleo, F., and Liverani, C.,
``Ergodic properties for a quantum nonlinear dynamics,''
 J. Stat. Phys. {\bf 97} (5/6), 957--1009 (1999).

\bibitem{JP}
Jak\v{s}i\'c, V., and Pillet, C.-A.,
``Ergodic properties of classical dissipative systems. I,''
 Acta Math. {\bf 181} (2), 245--282 (1998).

\bibitem{IL} Ibragimov, I.A., and Linnik, Yu.V.,
{\it Independent and Stationary Sequences of Random Variables},
edited by J. F. C. Kingman (Wolters--Noordhoff, Groningen, 1971).

\bibitem{Kat}
Katznelson, Y.,
{\it An Introduction in Harmonic Analysis}, 3rd edition (Cambridge University Press, 2004).

\bibitem{LL}
Lanford, III, O.E., and Lebowitz, J.L.,
Time Evolution and Ergodic Properties of Harmonic Systems, in
{\it Dynamical Systems, Theory and Applications},
 Lecture Notes in Physics, vol. 38 (Springer-Verlag, Berlin, 1975).

\bibitem{Na}
Nakazawa, H.,
``On the lattice thermal conduction,''
 Supplement of the Progress of Theor. Phys. {\bf 45}, 231--262 (1970).

\bibitem{P}
Peierls, R.E.,
 {\em Quantum Theory of Solids} (University Press, Oxford, 1956).

 \bibitem{RT}  Rey-Bellet, L., and Thomas, L.E.,
``Exponential convergence to non-equilibrium
stationary states in classical statistical mechanics,''
 Commun. Math. Phys. {\bf 225} (2), 305--329 (2002).

\bibitem{RLL}
Rieder, Z., Lebowitz,  J.L., and Lieb, E.,
``Properties of a harmonic crystal in a stationary nonequilibrium state,''
 J. Math. Phys. {\bf 8} (5), 1073--1078 (1967).

\bibitem{SL}
Spohn, H. and Lebowitz, J.L.,
``Stationary non equilibrium states of infinite harmonic systems,''
 Comm. Math. Phys. {\bf 54}, 97--120  (1977).


\bibitem{VF}
 Vishik, M.I. and Fursikov, A.V.,
{\it Mathematical Problems of Statistical Hydromechanics}
(Kluwer Academic, New York, 1988).

\bibitem{Lepri}
{\em Thermal Transport in Low Dimensions:
From Statistical Physics to Nanoscale Heat Transfer}, S.~Lepri (ed.),
Lecture Notes in Physics, vol. 921 (Springer, 2016).

\end{thebibliography}
\end{document}